\documentclass[a4paper,12pt]{article}
\pdfoutput=1
\usepackage{graphicx, rotating,amssymb,amsmath,multicol,soul,mathrsfs,eufrak}

\ifx\pdfoutput\undefined
\usepackage[dvips,bookmarks]{hyperref}	
\else
\usepackage{hyperref}	
\fi
\hypersetup{colorlinks,bookmarksopen,bookmarksnumbered,citecolor=verdec,
linkcolor=blue,pdfstartview=FitH,urlcolor=rossoc}
\def\myurl#1#2{\href{http://#1}{#2}}
\def\hhref#1{\href{http://arxiv.org/abs/#1}{#1}} 
\def\mhref#1{\href{mailto:#1}{#1}}		

\long\def\symbolfootnote[#1]#2{\begingroup\def\thefootnote{\fnsymbol{footnote}}\footnote[#1]{#2}\endgroup}

\usepackage{color}
\definecolor{rosso}{cmyk}{0,1,1,0.4}
\definecolor{rossos}{cmyk}{0,1,1,0.55}
\definecolor{rossoc}{cmyk}{0,1,1,0.2}
\definecolor{pink}{cmyk}{0,0.85,0.00,0.07}
\definecolor{blu}{cmyk}{1,1,0,0.3}
\definecolor{blus}{cmyk}{1,1,0,0.6}
\definecolor{blue}{cmyk}{1.0,0.30,0.2,0.1}
\definecolor{verde}{cmyk}{0.92,0,0.59,0.25}
\definecolor{verdec}{cmyk}{0.95,0,0.75,0.1}
\definecolor{verdes}{cmyk}{0.92,0,0.59,0.4}

\newcommand{\beq}{\begin{equation}}
\newcommand{\eeq}{\end{equation}}

\begin{document}

\begin{flushleft}
\scriptsize{ \hfill
SACLAY--T15/029}
\end{flushleft}

\color{black}
\vspace{0.3cm}

\begin{center}
{\Huge\bf PPPC 4 DM secondary: \\[1mm]
A Poor Particle Physicist Cookbook \\[0.5mm] for secondary radiation\\[2mm] from Dark Matter}

\medskip
\bigskip\color{black}\vspace{0.6cm}

{
{\large\bf Jatan Buch}\ $^{a,b}\, \symbolfootnote[1]{\mhref{jbuch.iitkgp@gmail.com}}$,
{\large\bf Marco Cirelli}\ $^{a}\, \symbolfootnote[2]{\mhref{marco.cirelli@cea.fr}}$, 
{\large\bf Ga\"elle Giesen}\ $^{a}\, \symbolfootnote[3]{\mhref{gaelle.giesen@cea.fr}}$,
{\large\bf Marco Taoso}\ $^{a}\, \symbolfootnote[4]{\mhref{marco.taoso@cea.fr}}$
}
\\[7mm]
{\it $^a$ \href{http://ipht.cea.fr/en/index.php}{Institut de Physique Th\'eorique}, Universit\'e Paris Saclay, CNRS, CEA,\\ F-91191 Gif-sur-Yvette, France}\\[3mm]
{\it $^b$ \href{http://www.iitkgp.ac.in/academics/?page=acadunits&&dept=MP}{Department of Physics}, Indian Institute of Technology, Kharagpur\\ West Bengal - 721302, India}\\[3mm]
\end{center}

\bigskip
 
\centerline{\large\bf Abstract}
\begin{quote}
\color{black}\large
We enlarge the set of recipes and ingredients at disposal of any poor particle physicist eager to cook up signatures from weak-scale Dark Matter models by computing two secondary emissions due to DM particles annihilating or decaying in the galactic halo, namely the radio signals from synchrotron emission and the gamma rays from bremsstrahlung. 
We consider several magnetic field configurations and propagation scenarios for electrons and positrons. 
We also provide an improved energy loss function for electrons and positrons in the Galaxy, including synchrotron losses in the different configurations, bremsstrahlung losses, ionization losses and Inverse Compton losses with an updated InterStellar Radiation Field. 

\end{quote}

\newpage
\tableofcontents

\bigskip

\section{Introduction}
\label{sec:introduction}

Dark Matter (DM) constitutes the largest matter component of the Universe, but its nature has so far remained elusive. It is searched for in a number of ways, and most notably via its possible electromagnetic emission in the Galaxy. 

\medskip

In particular, an interesting strategy consists in looking for the emissions produced by the interactions of relativistic electrons and positrons, injected by DM annihilations (or decays), with the galactic environment. These emissions go under the collective name of `secondary radiation' and are essentially of three kinds: (i) radio waves due to the synchrotron radiation of the $e^\pm$ on the galactic magnetic field; (ii) gamma rays due to the bremsstrahlung processes on the galactic gas density; (iii) gamma rays due to the Inverse Compton scattering (ICS) processes on the interstellar radiation field.

\medskip

They have received different attention in the literature. Synchrotron emission has been considered since a long time in regions close to the Galactic Center (GC), characterized by a large intensity of the magnetic field~\cite{radioGC}. Its relevance in wider regions of interest in the Galaxy has also been highlighted~\cite{Blasi:2002ct,Fornengo:2011iq,Borriello:2008gy,Borriello:2008dt,Delahaye:2011jn,Linden:2011au,Mambrini:2012ue,Fornengo:2011cn,Fornengo:2011xk,Hooper:2012jc,Carlson:2012qc}.
Bremsstrahlung gamma rays have mostly been neglected for what concerns DM studies. Recently, however, their importance has been recognized, especially in connection with searches for relatively light (10-40 GeV) DM signals from the GC~\cite{Cirelli:2013mqa,Abazajian:2014fta,Daylan:2014rsa,Lacroix:2014eea,Cirelli:2014lwa,Abazajian:2014hsa}. Finally, ICS gamma rays have been identified as an important component of the DM gamma ray spectrum mainly in conjunction with the models of leptophilic DM featuring a large annihilation cross section, proposed in the wake of the lepton excesses measured by {\sc Pamela, Fermi, Hess} and most recently {\sc Ams-02} (see e.g.~\cite{Eichler:2005aa,Cholis:2008wq,Zhang:2008tb,Borriello:2009fa,Barger:2009yt,Cirelli:2009vg} and many subsequent works).

\medskip

All these signatures are  potentially very relevant and promising. Their practical use, however, depends on a number of different choices, e.g.~related to the unknown magnetic field configuration, to the unknown propagation parameters of electrons and positrons in the Galaxy, to the unknown gas distribution, to the unknown profile of Dark Matter etc. Some of these uncertainties are also shared with other possible signals from DM in other Indirect Detection channels. 
In pursuing the goal of identifying Dark Matter or better constraining it, it is therefore important to be able to perform multi-messenger analyses in a coherent framework and therefore to develop a set of coherent, model independent tools. A step in this direction is what we would like to make here.

\medskip

More precisely, the purpose of this paper is to provide state-of-the-art tools allowing the computation of synchrotron and bremsstrahlung radiation~\footnote{We remind that the equivalent tool for ICS radiation is already provided in~\cite{PPPC4DMID}, and will be updated soon in order to include the improvements of the present paper.} for any (weak-scale) DM model, for a set of possible astrophysical configurations that bracket the current sensible ranges of the uncertainties. This follows the spirit of previous papers (\cite{PPPC4DMID,PPPC4DMnu}): the concrete goal is to enable the `DM model builder' to readily compute the synchrotron and bremsstrahlung phenomenology of her model without having to fiddle with the underlying computations or even with the intervening astrophysics, but just by choosing which configurations to adopt (and being able to adopt the same choices consistently for other indirect detection channels).
On the way to achieve that, we have to upgrade some ingredients used to accurately compute the population of DM-induced electrons and positrons, namely the energy loss function and the $e^\pm$ halo functions (see below).
In the spirit of \cite{PPPC4DMID,PPPC4DMnu} we always employ semi-analytical methods rather than fully numerical ones, in order to better control the relevant physics.

Armed with the tools just described, we will later play the same game ourselves~\cite{future_synchr} by applying them to refine the constraints from synchrotron radiation derived in~\cite{Fornengo:2011iq}. 

\medskip

The rest of this paper is organized as follows. 
In Sec.~\ref{sec:astro configurations} we spell out the astrophysical ingredients we need. We discuss in some detail the configurations of the magnetic field (relevant for synchrotron emission) while we just recall the main points concerning the other ingredients (DM distribution, CR propagation parameters, ISRF, galactic gas maps).
In Sec.~\ref{sec:results} we list several new results. 
First, in Sec.~\ref{sec:energy loss} and~\ref{sec:halo functions} we present our improved energy loss function and our improved $e^\pm$ generalized halo functions.
Then, in Sec.~\ref{sec:synchr formalism} and~\ref{sec:brem formalism} we present the two main results and numerical outputs of this work: the generalized halo functions for synchrotron radiation and for bremsstrahlung emission.
In Sec.~\ref{sec:conclusions} we conclude.

\section{Astrophysical configurations}
\label{sec:astro configurations}
In this Section we spell out the astrophysical ingredients that we use to compute the propagation of electrons and positrons and ultimately the synchrotron and bremsstrahlung signals. 

\medskip

While there certainly exist some interdependences between the parameters entering in these astrophysical ingredients (for instance, the thickness of the CR diffusive halo --called $L$ in the notation below-- is related to the vertical extent of the galactic magnetic field, since a far reaching magnetic field determines a thick CR confinement box) it is beyond our scope to impose such correlations by hand. Our aim is to provide the full range of possible choices and it is up to the user to choose sensible combinations. In order to consider inter-dependencies self-consistently one is better off choosing a fully numerical approach to the propagation of cosmic rays, such as by using {\tt Galprop}~\cite{galprop} or {\tt Dragon}~\cite{dragon}.

\subsection{Galactic Magnetic Field}
\label{sec:magnetic field}

Our Galaxy has a complicated magnetic field structure, and dedicated efforts by several groups have been performed in order to map it: for some recent overviews and sets of references see for instance~\cite{Beck:2011he,Jansson:2009ip,Pshirkov:2011um,Beck:2013bxa}. We recall here the salient features of the inferred magnetic field and then define the simplified functional form that we will adopt. 

\medskip

The total galactic magnetic field $\vec{B}_{\rm tot}$ is the sum of a large scale regular $\vec{B}_{\rm reg}$ and a turbulent $\vec{B}_{\rm turb}$ component. These, in turn, can be decomposed in different contributions, including disk and halo fields. 
The regular magnetic field is caused by dynamo effects in the galaxy and it can be studied with Faraday rotation measurements of nearby pulsars and high latitude radio sources, or with measurements of the polarized synchrotron intensity. On the other hand, the turbulent magnetic fields are  tangled by turbulent gas flows and can be traced looking for their unpolarized synchrotron emission.
Recent models of the galactic magnetic fields have been proposed e.g. in~\cite{Sun2008,Jansson:2012rt,Jansson:2012pc,Sun:2010sm,     Pshirkov:2011um}.

\medskip

Rather than the detailed magnetic field geography, the overall intensity is more important for our purposes.
While we keep in mind that the complicated cartography sketched above can have an impact on the determination of the energy losses of electrons and on the synchrotron emission from DM, we choose to model the disk field strength by a double exponential in $z$ and  in $r$, as proposed e.g.~by \cite{Strong:1998fr} and \cite{Han2006} for the radial part. Namely, we use
\begin{equation}
B_{\rm tot}= B_0 \exp \left( -\frac{r-r_{\odot}}{R_D}- \frac{\left|z\right|}{z_D}\right)
\end{equation}
where $r_\odot = 8.33$ kpc is the location of the Sun. We then adopt several configurations for the values for the parameters $B_0$, $r_D$ and $z_D$, as shown in the table in fig.~\ref{fig:astroparam}:
\begin{itemize}
\item[{\scriptsize $\blacksquare$}] Model 1 (MF1 for ``Magnetic Field 1'' hereafter) is the configuration used in~\cite{PPPC4DMID} and very similar to the one used in the original {\sc Galprop} code (it differs by the normalization factor $B_0$, which has changed a few times in the {\sc Galprop} literature~\cite{Strong:1998fr,Strong:1998pw,Moskalenko:1997gh}). 
\item[{\scriptsize $\blacksquare$}] Model 2 (``MF2") is loosely based on the findings of~\cite{Sun2008} (and previous~\cite{Han2006}). Following one of the models in~\cite{Sun2008} we take a value of 2.1 $\mu G$ for the intensity of the disk regular field at solar location (we report it to our value for $r_\odot$); we then add an intensity of 3  $\mu G$ to account for the random component.  The resulting field is steeper in $r$ and in $z$ than MF1 and reaches slightly higher values at the GC. 
\item[{\scriptsize $\blacksquare$}] Model 3 (``MF3") is modeled following~\cite{Strong:2011wd}. It is substantially higher at the location of the Earth and has larger scale heights both in $r$ and in $z$, i.e. it extends much farther out in both directions.
\end{itemize}

\subsection{Other astrophysical ingredients}
\label{sec:other astro}

\begin{figure}[p]
\begin{minipage}{0.5\textwidth}
\vspace{-0.2cm}
\centering
\includegraphics[width=\textwidth]{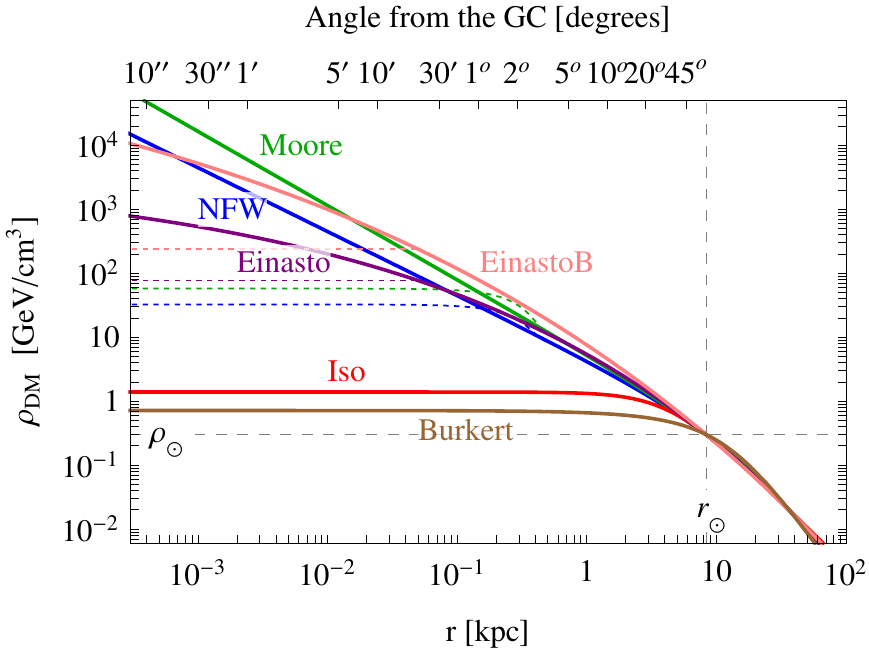}
\end{minipage}
\quad
\begin{minipage}{0.4\textwidth}
\vspace{-0.5cm}
\centering
\footnotesize{  
\begin{tabular}{l|crc}
\multicolumn{4}{l}{\bf Dark Matter halo profiles}\\
\hline\\[-3mm]
Halo & $\alpha$ &  $r_{s}$ [kpc] & $\rho_{s}$ [GeV/cm$^{3}$]\\
  \hline \\[-3mm]
  NFW & $-$ & 24.42 & 0.184 \\
  Einasto & 0.17 & 28.44 & 0.033 \\
  EinastoB & 0.11 & 35.24 & 0.021 \\
  Isothermal & $-$ & 4.38 & 1.387 \\
  Burkert & $-$ & 12.67 & 0.712 \\
  Moore & $-$ & 30.28 & 0.105 
 \end{tabular}} 
\vspace{0.3cm} 
\footnotesize{  
\begin{tabular}{c|ccc}
\multicolumn{4}{l}{\bf Propagation parameters for $e^\pm$}\\
\hline\\[-3mm]
Model  & $\delta$ & $\mathcal{K}_0$ [kpc$^2$/Myr] & $L$ [kpc]  \\
\hline 
{\sc Min}  & 0.55 & 0.00595 & 1 \\
{\sc Med} & 0.70 & 0.0112 & 4  \\
{\sc Max} & 0.46 & 0.0765 & 15 
\end{tabular}} 
\end{minipage}

\vspace{4mm}

\begin{minipage}{0.39\textwidth}
\vspace{-0.8cm}
\footnotesize{  
\begin{tabular}{l|c|c|c|c}
\multicolumn{5}{l}{\bf Magnetic field configurations}\\
\hline\\[-3mm]
Model & ref.& $B_0$ & $r_D$ & $z_D$ \\
 &       & [$\mu G$] & [kpc] & [kpc] \\
\hline \\[-4.mm]
MF1& \cite{Strong:1998fr} & 4.78& 10& 2 \\
MF2& \cite{Sun2008,Han2006}&5.1& 8.5& 1\\
MF3& \cite{Strong:2011wd} &9.5& 30& 4\\
\end{tabular}}
\end{minipage}
\quad
\begin{minipage}{0.60\textwidth}
\flushright
\includegraphics[width= 0.47 \textwidth]{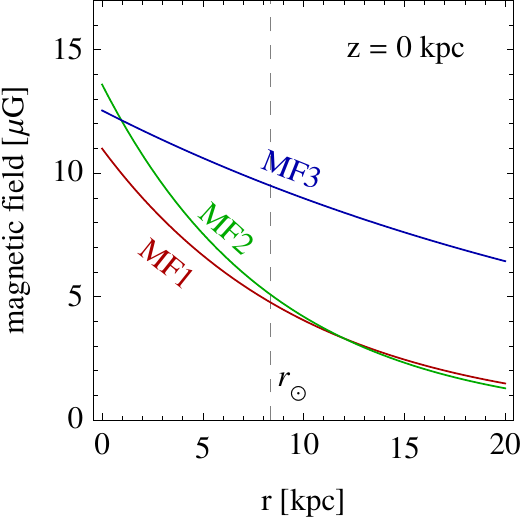}\quad
\includegraphics[width= 0.47 \textwidth]{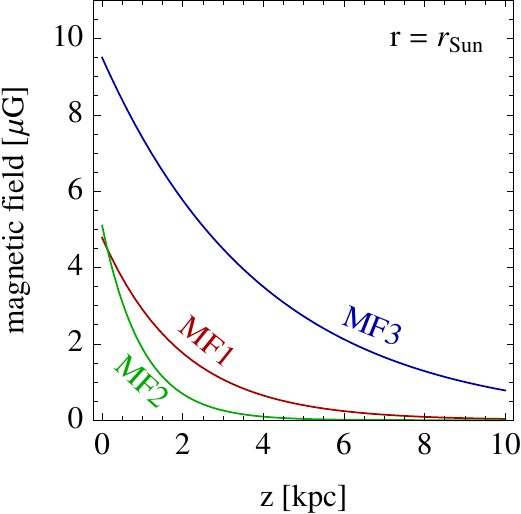}
\end{minipage}

\vspace{0.2cm}

\begin{minipage}{1\textwidth}
\centering
\includegraphics[width=0.475 \textwidth]{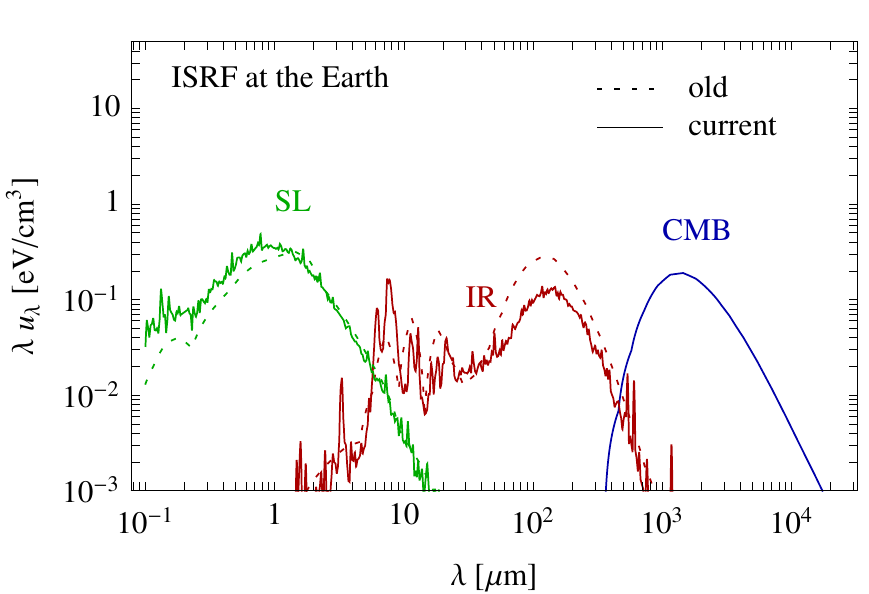} \hfill
\includegraphics[width=0.47 \textwidth]{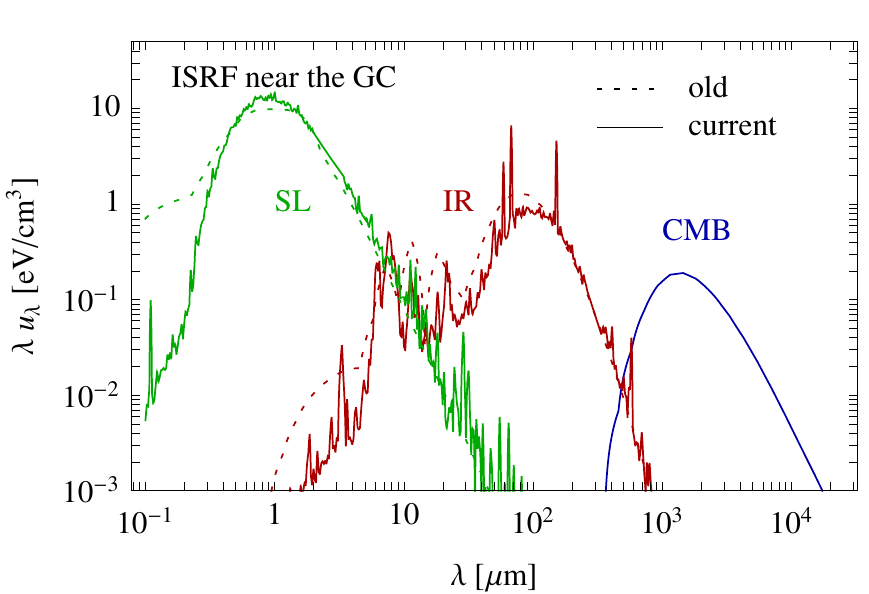}
\end{minipage}

\vspace{3mm}

\begin{minipage}{1\textwidth}
\includegraphics[width=0.31\textwidth]{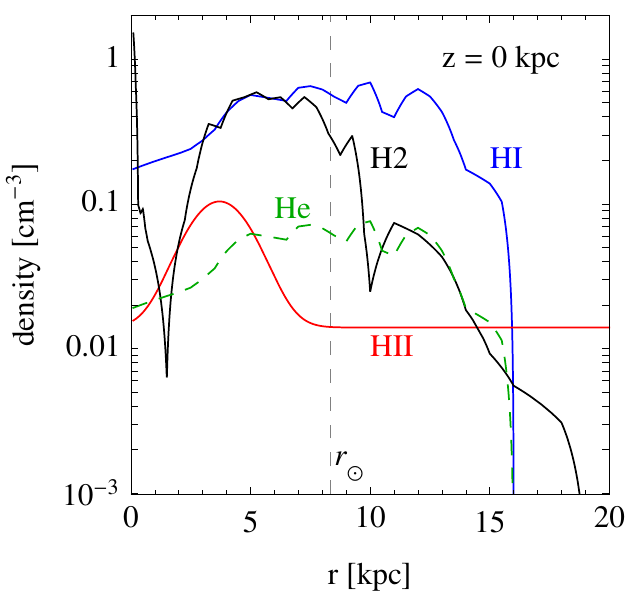}\quad
\includegraphics[width=0.31\textwidth]{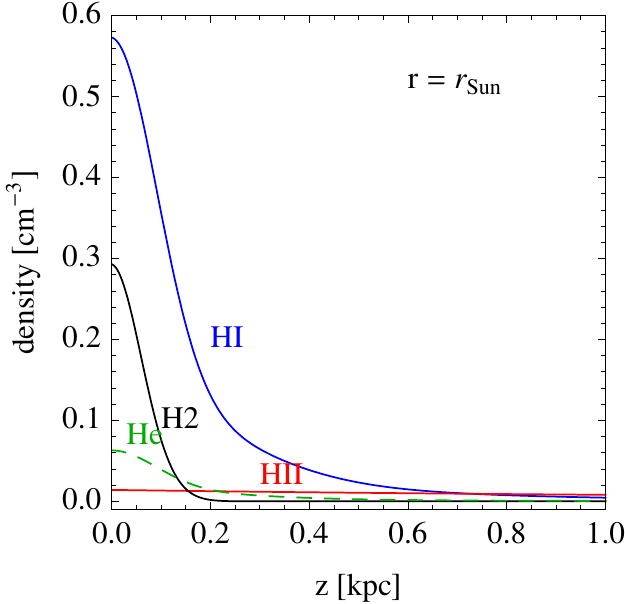}\quad
\includegraphics[width=0.31\textwidth]{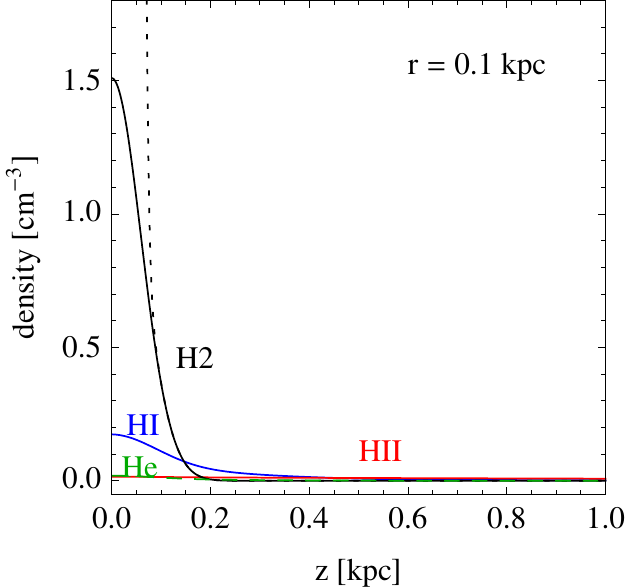}
\end{minipage}

\vspace{0mm}
\caption{\em \small \label{fig:astroparam} Collection of the {\bfseries astrophysical ingredients} we use. Top row: DM profiles (figure taken from~\cite{PPPC4DMID}) and propagation parameters for electrons and positrons in the Galaxy. Second row: magnetic field configurations. Third row: illustration of the ISRF in two sample locations. Bottom row: illustration of the galactic gas densities (figure from~\cite{Cirelli:2013mqa}).} 
\end{figure}

In this section we recall the other astrophysical ingredients involved in the computations. We illustrate most of them in Fig.~\ref{fig:astroparam}. As a general rule, we want to use state-of-the-art but standard ingredients, in order to allow easy comparison with other work. 

\begin{itemize}
\item[$\circ$] The {\bf DM density profile} in the Galaxy. We adopt the 6 standard profiles as defined in~\cite{PPPC4DMID} (to which we refer for references and some discussion). They always assume spherical symmetry and their functional forms as functions of the galactocentric coordinate\footnote{We distinguish the notation between the galactocentric coordinate ${\mathbf r}$ and the cylindrical coordinates $(r,z)$ that we will use in most of the following. Obviously ${\mathbf r} = \sqrt{r^2+z^2}$.}  ${\mathbf r}$ read:
\begin{equation}
\begin{array}{rrcl}
{\rm NFW:} & \rho_{\rm NFW}({\mathbf r})  & = & \displaystyle \rho_{s}\frac{r_{s}}{{\mathbf r}}\left(1+\frac{{\mathbf r}}{r_{s}}\right)^{-2} \\[4mm]
{\rm Einasto:} & \rho_{\rm Ein}({\mathbf r})  & = & \displaystyle \rho_{s}\exp\left\{-\frac{2}{\alpha}\left[\left(\frac{{\mathbf r}}{r_{s}}\right)^{\alpha}-1\right]\right\} \\[4mm]
{\rm Isothermal:} &  \rho_{\rm Iso}({\mathbf r}) & = & \displaystyle \frac{\rho_{s}}{1+\left({\mathbf r}/r_{s}\right)^{2}} \\[4mm]
{\rm Burkert:} & \rho_{\rm Bur}({\mathbf r}) & = & \displaystyle  \frac{\rho_{s}}{(1+{\mathbf r}/r_{s})(1+ ({\mathbf r}/r_{s})^{2})} \\[4mm]
{\rm Moore:} &  \rho_{\rm Moo}({\mathbf r})  & = & \displaystyle \rho_{s} \left(\frac{r_s}{{\mathbf r}}\right)^{1.16} \left(1+\frac{{\mathbf r}}{r_s}\right)^{-1.84} .
\end{array}
\label{eq:profiles}
\end{equation} 
The profiles are plotted in the top left panel of Fig.~\ref{fig:astroparam} and their parameters reported in the corresponding Table. We remind here that they are normalized by requiring that the density at the location of the Sun $r_\odot = 8.33$ kpc be 0.3 GeV/cm$^3$ and the total mass of the Milky Way be 4.7 $\times 10^{11} \ M_\odot$. Satisfying these two criteria allows to fix the $r_s$ and $\rho_s$ parameters. Other normalizations (e.g. of the density at the Sun) are used in the literature and they would modify the profiles. 

\item[$\circ$] Electron and positron {\bf propagation parameters}. These have to be plugged in the diffusion-loss equation that we will discuss in detail in Sec.~\ref{sec:halo functions}. We adopt the standard choices {\sc Min}, {\sc Med}, {\sc Max} as reported in the table in Fig.~\ref{fig:astroparam}. Here $\delta$ and ${\mathcal K}_0$ are the exponent of the energy dependence of the diffusion coefficient and its normalization while $L$ is the thickness of the diffusive halo. While an updated assessment of the validity of these ranges of parameters would be welcome, especially in the light of the wealth of recent data\footnote{For instance, a string of recent papers, based precisely on synchrotron radio emission~\cite{DiBernardo:2012zu,Bringmann:2011py,Orlando:2013ysa,Fornengo:2014mna} but also on positrons~\cite{DiMauro:2014iia,Lavalle:2014kca} and somewhat also on gamma rays~\cite{Fermidiffuse} and antiprotons~\cite{Giesen:2015ufa}, finds that the thin halo predicted by {\sc Min} is seriously disfavored.}, this would be beyond our current scope. We continue using the standard values reported here also for consistency reasons.

\item[$\circ$] {\bf InterStellar Radiation Field} (ISRF). Electrons and positrons propagating in the galactic halo lose energy by Inverse Compton scattering on the ambient light. A detailed description of this radiation field is therefore important in order to reliably compute the energy losses. We adopt the latest radiation maps extracted from {\sc Galprop}~\cite{ISRFGalprop}. These replace the ones formerly used in the literature, and in particular in~\cite{PPPC4DMID}. In fig.~\ref{fig:astroparam} we draw the two maps in two sample locations (at the Earth and near the galactic center) and compare them. One clearly sees the three different components (StarLight SL, InfraRed IR and the CMB blackbody spectrum). The current map is much more detailed and normalization differences of the order of a factor 2 are visible, but the overall behavior is confirmed. We will see that these small differences have an (equally small) impact on the observables entering $e^\pm$ propagation.

\item[$\circ$] {\bf Gas maps}. Electrons and positrons also lose energy by processes occurring on the interstellar atomic and molecular gas (Coulomb interactions, ionization, bremsstrahlung). We use the gas maps described in~\cite{Moskalenko:2001ya} and already used in~\cite{Cirelli:2013mqa}. We refer to the latter for some discussion. They are illustrated in fig.~\ref{fig:astroparam}. The relevant species are atomic (HI) and molecular (H$_2$) neutral hydrogen, ionized hydrogen (HII), neutral atomic helium (He) and ionized helium (which is however irrelevant for all practical purposes). As discussed in particular in~\cite{Cirelli:2013mqa}, these maps represent a reliable description of the coarse grained distribution of gas in the Galaxy, but miss important features at small scales. In particular, they do not take into account the regions characterized by a much higher gas density (up to 2 or 3 orders of magnitude with respect to the coarse grained maps) which are known to exist close to the galactic center (typically at ${\mathbf r} \lesssim 200$ pc scales). For the purpose of the general tools that we are developing in this work, we do not correct by hand the coarse-grid maps by adding the high density regions (contrary to what was done in~\cite{Cirelli:2013mqa}) but we will allow the user to change the overall normalization of the gas density in the energy loss function that we will describe below.
\end{itemize}


\section{Results}
\label{sec:results}

\subsection{An improved energy loss function for $e^\pm$ in the Galaxy}
\label{sec:energy loss}

\begin{figure}[t]
\begin{center}
\hspace{-1.05cm}
\includegraphics[width= 0.545 \textwidth]{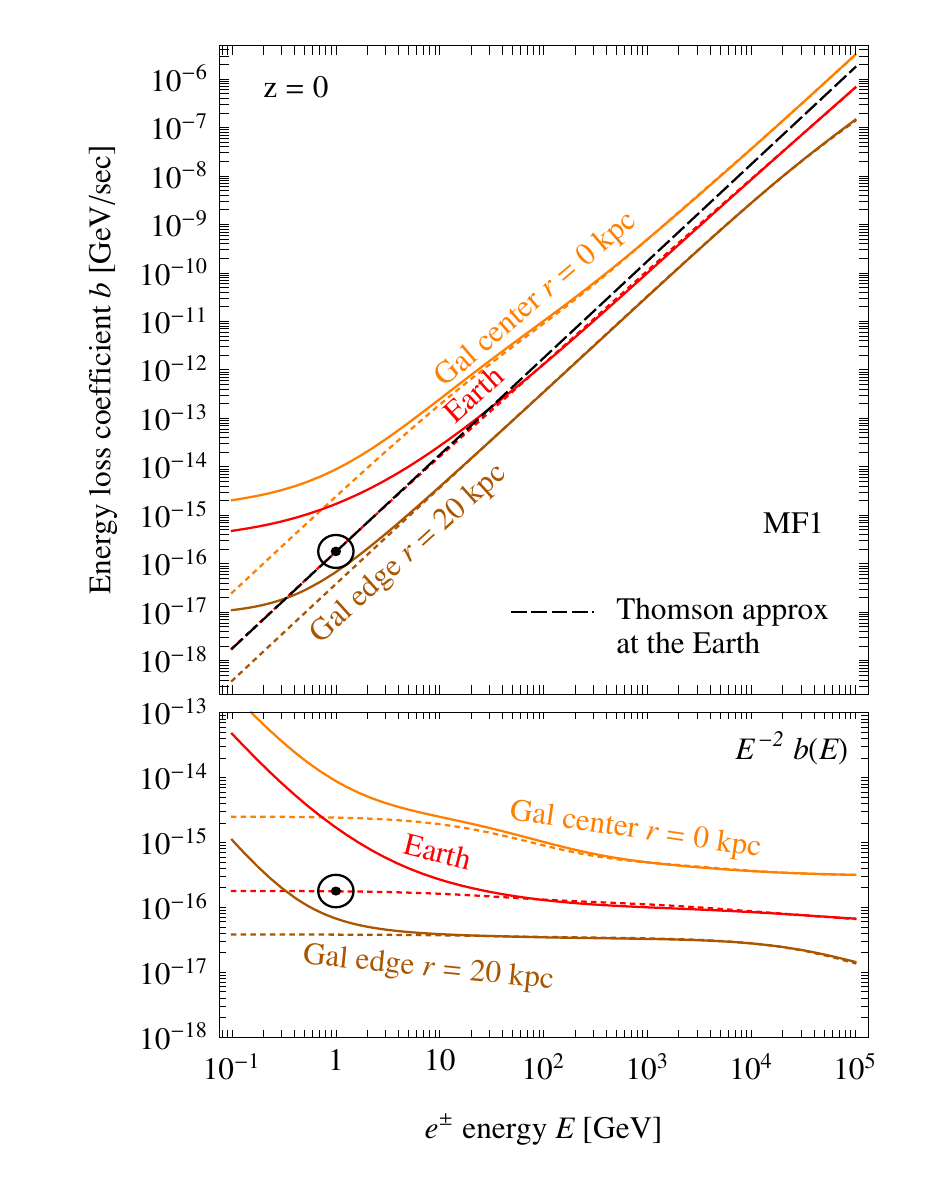}
\hspace{-0.90cm}
 \includegraphics[width= 0.545 \textwidth]{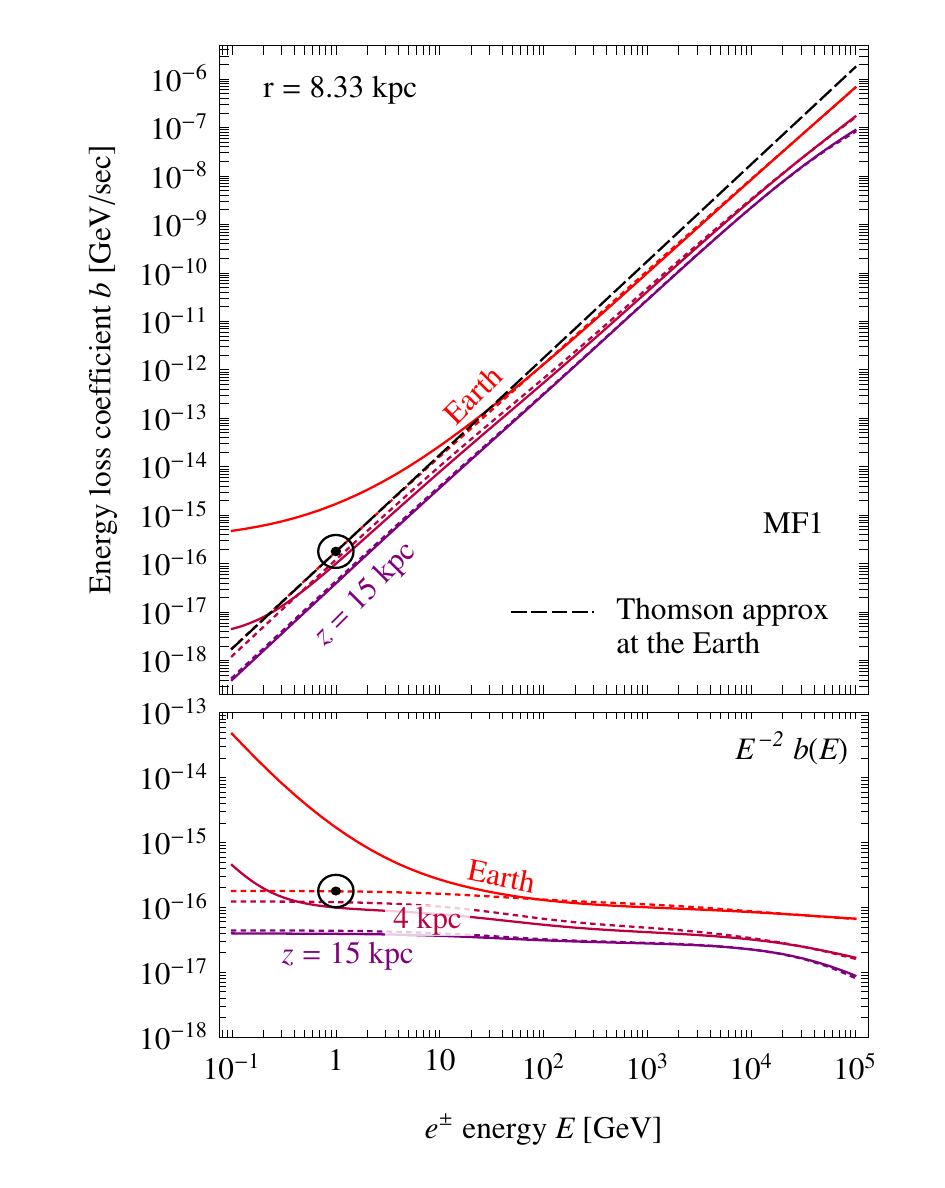}
\caption{\em \small \label{fig:b} {\bfseries Energy loss  function} for electrons and positrons in the Milky Way. Left panel: in the galactic disk ($z=0$), at several locations along the radial coordinate $r$. Right panel: above (or below) the location of the Earth along the coordinate $z$. Here the magnetic field model MF1 has been fixed for definiteness. The circled dot identifies the constant value sometimes adopted. The dotted colored lines are the same function before the improvements listed in Sec.~\ref{sec:energy loss}. This figure replaces the analogous one (fig.~5) of~\cite{PPPC4DMID}.}
\end{center}
\end{figure}

\begin{figure}[t]
\begin{center}
\includegraphics[width=0.50\textwidth]{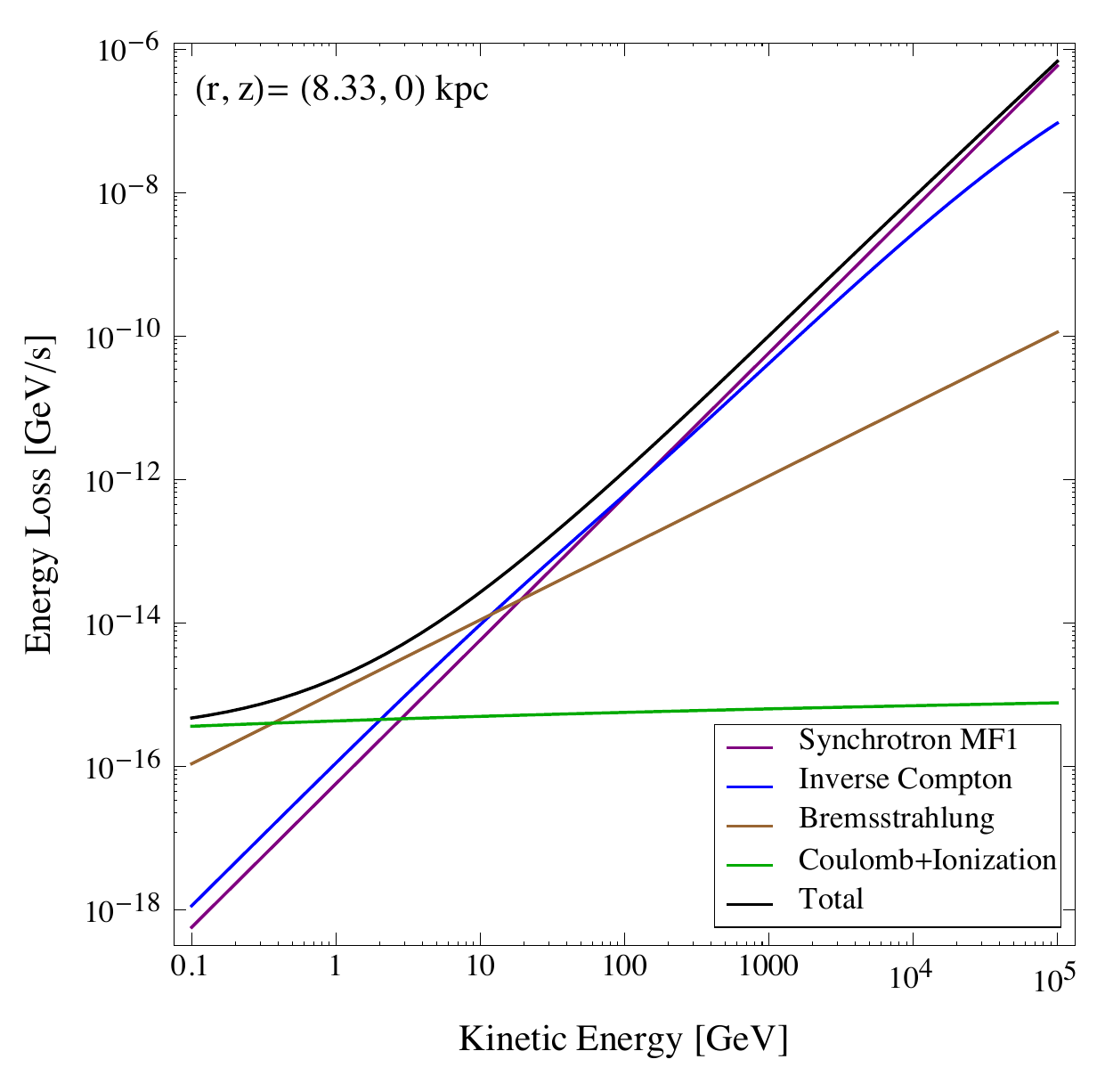} 
\hspace{-0.10cm}
\includegraphics[width=0.48\textwidth]{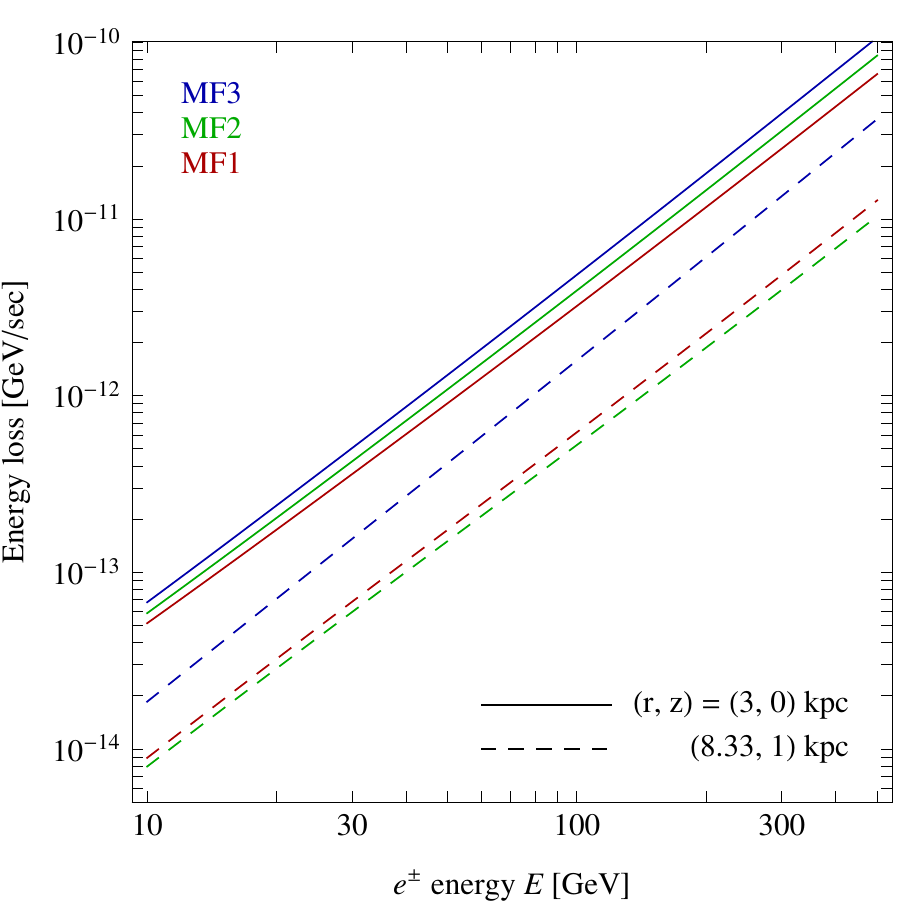}
\caption{\small \em\label{fig:SynchLoss} Left panel: the {\bfseries different processes} contributing to the energy loss function, at the location of the Earth. Right panel: the dependence of the energy loss coefficient function on the {\bfseries choice of magnetic field model}, in two locations. 
}
\end{center}
\end{figure}

Using the ingredients described above, we compute an improved function describing the energy losses of electrons and positrons during their propagation in the Galaxy. It includes energy losses by Coulomb interactions with the interstellar gas, by ionization of the same gas, by bremsstrahlung on the same gas, by ICS (using the updated ISRF presented in Sec.~\ref{sec:other astro}) and by synchrotron emission, with the choice of the three magnetic field models discussed in Sec.~\ref{sec:magnetic field}. Schematically: 
\begin{equation}
\label{eq:btot}
b_{\rm tot}(E,r,z) \equiv -\frac{{\rm d}E}{{\rm d}t} = b_{\rm Coul+ioniz} + b_{\rm brem} + b_{\rm ICS} + b_{\rm syn}
\end{equation}
where $E$ is the energy of the electron or positron and $r$ and $z$ are cylindrical galactic coordinates. 
Such a function is provided on the~\myurl{www.marcocirelli.net/PPPC4DMID.html}{website}~\cite{website} in the format \ {\tt btot[E,r,z,gasnorm,MF]}, where {\tt gasnorm} allows to change the overall normalization of the gas densities and {\tt MF} is a flag selecting the magnetic field model. We now recall the different components of this function~\footnote{Notice that we will always limit ourselves to the case of relativistic electrons.} and illustrate its main features in fig.~\ref{fig:b} and~\ref{fig:SynchLoss}. Details can be found in standard references such as~\cite{Blumenthal:1970gc,Schlickeiser} as well as in~\cite{PPPC4DMID,Cirelli:2009vg}.

\medskip

\begin{itemize}
\item {\bf Energy losses by Coulomb interaction and ionization} on neutral matter are described by 
\begin{equation}
\label{eq:enlossCIneut}
b^{\rm neut}(E,r,z) = \frac94\ c\ \sigma_{\rm T}\, m_e \sum_i n_i Z_i \left( \log \frac{E}{m_e} +\frac23 \log\frac{m_e}{\Delta E_i}\right)
\end{equation}
where $c$ is the speed of light, $\sigma_T = 8\pi  r_e^2/3$, with $r_e = \alpha_{\rm em}/m_e$, is the Thompson cross section, $n_i$ is the number density of gas species $i$ with atomic number $Z_i$ and $\Delta E_i$ is its average excitation energy (it equals 15 eV for hydrogen and 41.5 eV for helium).

On ionized matter, one has 
\begin{equation}
\label{eq:enlossCIion}
b^{\rm ion}(E,r,z) = \frac34\ c\ \sigma_{\rm T}\, m_e\ n_e \left( \log \frac{E}{m_e} +2 \log\frac{m_e}{E_{\rm pla}}\right)
\end{equation}
where $n_e$ is the electron density and $E_{\rm pla} = \sqrt{4 \pi \, n_e\, r_e^3}\, m_e/\alpha$ corresponds to the characteristic energy of the plasma. 

The total energy losses for Coulomb interactions and ionization processes, $b_{\rm Coul+ioniz} = b^{\rm neut}+ b^{\rm ion}$, will therefore be given by the sum of eq.~(\ref{eq:enlossCIneut}) and eq.~(\ref{eq:enlossCIion}) with, respectively, the densities of ionized and neutral gas species. In both cases, energy losses are {\em essentially independent} of $E$, since the constant terms in the brackets are numerically dominant.

\item {\bf Energy losses by bremsstrahlung} are described by
\begin{equation}
\label{eq:enloss}
b_{\rm brem}(E,r,z)  = c\,\sum_i n_i(r,z)\int_0^{E} {\rm d} E_\gamma\,E_\gamma\,\frac{{\rm d} \sigma_i}{{\rm d} E_\gamma}\,,
\end{equation}
where $E_\gamma$ corresponds to the energy of the gamma ray emitted in each bremsstrahlung process. The differential cross-section reads
\begin{equation}
\label{eq:sigma}
\frac{{\rm d} \sigma_i(E,E_\gamma)}{{\rm d} E_\gamma}=\frac{3\,\alpha_{\rm em}\sigma_T}{8\pi\,E_\gamma}\left\{\left[1+\left(1-\frac{E_\gamma}{E}\right)^2\right]\phi^i_1-\frac{2}{3}\left(1-\frac{E_\gamma}{E}\right)\phi^i_2\right\}\,,
\end{equation}
where $\phi^i_{1,2}$ are scattering functions dependent on the properties of the scattering system. 

For a completely ionized gas plasma with charge $Z$ one has
\begin{equation}
\label{eq:phiplasma}
\phi^{\rm ion}_1(E,E_\gamma)=\phi^{\rm ion}_2(E,E_\gamma)=4(Z^2+Z)\left\{\log\left[\frac{2E}{m_e\,c^2}\left(\frac{E-E_\gamma}{E_\gamma}\right)\right]-\frac{1}{2}\right\}\,,
\end{equation}
and thus the energy losses in this regime (`weak shielding') read
\begin{equation}
\label{eq:enlossWS}
b_{\rm brem}^{\rm ion} = \frac{3}{2\pi}c\, \alpha_{\rm em} \,  \sigma_{\rm T} \, n_i \ Z(Z+1) \left(\log \left( 2 \ \frac{E}{m_e}\right)-\frac{1}{3} \right) E\,.
\end{equation}

On the other hand, for atomic neutral matter the scattering functions have a more complicated dependence, which is usually parameterized in terms of the  quantity
$\Delta = \frac{E_\gamma m_e}{4 \alpha_{\rm em} E (E-E_\gamma)}$. For the relativistic regime we are interested in, since $E \gtrsim 1$ MeV always, one basically cares for the limit $\Delta \to 0$ for which these functions are constant and take the following numerical values: 
\begin{equation}
\label{eq:phiSS}
\begin{aligned}
\phi^{\rm H}_{1}(\Delta = 0) & \equiv \phi^{\rm H}_{1, {\rm ss}} = 45.79,\\ 
\phi^{\rm H}_{2}(\Delta = 0) & \equiv \phi^{\rm H}_{2, {\rm ss}} = 44.46, \\
\phi^{\rm He}_{1}(\Delta = 0) & \equiv \phi^{\rm He}_{1, {\rm ss}} = 134.60, \\
\phi^{\rm He}_{2}(\Delta = 0) & \equiv \phi^{\rm He}_{2, {\rm ss}} = 131.40, \\
\phi^{{\rm H}_2}_{(1,2)}(\Delta = 0) &\simeq 2\, \phi^{\rm H}_{(1,2), {\rm ss}}.
\end{aligned}
\end{equation} 
The subscript $_{\rm ss}$ in this notation refers to the fact that this regime is usually called `strong-shielding' because the atomic nucleus is screened by the bound electrons and the impinging $e^\pm$ have to force the shield. In this limit the energy losses read
\begin{equation}
\label{eq:enlossSS}
b_{\rm brem}^{\rm neut} = \frac{3}{8\pi}c\, \alpha_{\rm em} \, \sigma_{\rm T} \,  n_i \left(\frac{4}{3} \phi^i_{1,{\rm ss}} -\frac{1}{3} \phi^i_{2,{\rm ss}} \right) E\,.
\end{equation}

The total energy losses for bremsstrahlung will therefore be given by the sum of eq.~(\ref{eq:enlossWS}) and eq.~(\ref{eq:enlossSS}) with, respectively, the densities of ionized and neutral gas species. In both cases, at leading order, energy losses are {\em linearly dependent} on $E$. A further logarithmic dependence arises for scattering in ionized medium, while a small additional energy dependence is also found in neutral medium if one accounts for the effect of finite $\Delta$. 

\item {\bf Energy losses by Inverse Compton Scattering} are described, in exact form, by
\begin{equation}
\begin{split}
& b_{\rm ICS} = \\
& 3c\, \sigma_{\rm T} \int_0^\infty d\epsilon\, \epsilon \int_{1/4\gamma^2}^1dq\ n(\epsilon) \frac{(4\gamma^2-\Gamma_\epsilon)q-1}{(1+\Gamma_\epsilon q)^3}\left[ 2q\ln q+q+1-2q^2+\frac{1}{2}\frac{(\Gamma_\epsilon q)^2}{1+\Gamma_\epsilon q}(1-q) \right], 
\end{split}
\label{eq:enlossICS}
\end{equation}
where $n(\epsilon,r,z)$ is the number density (per unit volume and unit energy) of photons of the ISRF, with energy $\epsilon$, $\gamma = E/m_e$ is the relativistic factor of the electrons and positrons and $\Gamma_\epsilon = 4\epsilon\gamma/m_e$.

In the Thomson limit (valid for low electron energies), they reduce to the particularly compact expression
\beq
b_{\rm ICS} = \frac{4\,c\ \sigma_{\rm T}}{3\, m_e^2}\, E^2  \int_0^\infty d\epsilon\ \epsilon\ n(\epsilon,r,z) \qquad {\rm [Thomson\ limit]},
\label{eq:enlossICSThomson}
\eeq
which makes the energy density in the photon bath $u_{\rm ISRF} = \int d\epsilon\, \epsilon\, n(\epsilon,r,z)$ apparent.

The ICS energy losses are {\it proportional to} $E^2$ (as evident in the Thomson expression, but also in eq.~(\ref{eq:enlossICS}) noting that $4\gamma^2 q$ is the dominant piece at the numerator) for small $E$. For large $E$, the dependence softens. 

\item {\bf Energy losses by synchrotron emission} are described by
\begin{equation}
b_{\rm syn} = \frac{4\,  c \ \sigma_{\rm T}}{3\, m_e^2}\, E^2\, \frac{B^2}{8\pi}
\label{eq:enlosssyn}
\end{equation}
where $B$ is the strength of the magnetic field. This formula is in close analogy to the one for ICS losses: the integral term in~(\ref{eq:enlossICSThomson}) and the $B^2$ term in~(\ref{eq:enlosssyn}) correspond to the energy density in the photon bath and in the magnetic field respectively. In particular, synchrotron energy losses are also {\it proportional to} $E^2$.

\end{itemize}

\medskip

\noindent 
In fig.~\ref{fig:SynchLoss}, left panel, we plot the different energy losses discussed above, at the location of the Earth. The different dependences on the $e^\pm$ energy are clearly shown. Hence, the dominant process in the different energy ranges are, in order, ionization (including Coulomb), bremsstrahlung, ICS and synchrotron.

\medskip 

In fig.~\ref{fig:b} we plot the total energy loss function in several locations in the galactic plane (left panel) and at several galactic altitudes at the location of the Earth (right panel). We compare it with the previous version of the same function not including the improvements listed at the beginning of this section (dashed colored lines). The main modification is apparent at low energies and it is due to the inclusion of bremsstrahlung, ionization and Coulomb losses. Being related to the presence of gas, it disappears at the locations outside of the galactic disk. 

The modifications due to the use of the new ISRF is minimal and mostly concentrated at low energies, so it is hidden by the dominant bremsstrahlung, ionization and Coulomb losses in most cases except well outside of the plane where the absence of gas makes it indeed visible (see the slight difference between the solid and dashed purple lines corresponding to $z = 15$ kpc in the right panel).

\medskip 

While in fig.~\ref{fig:SynchLoss} left and in fig.~\ref{fig:b} we have chosen the MF1 for definiteness, in fig.~\ref{fig:SynchLoss} right we explore the impact of changing the magnetic field model. Not surprisingly, in $(r,z)=(3,0)$ kpc the synchrotron energy losses are larger than at $(r,z)=(8.33,1)$ kpc, and the ordering reflects the intensity of the magnetic field in the corresponding model (see fig.~\ref{fig:astroparam}).

\medskip 

In the next subsection we employ this improved energy loss function to compute the halo functions for electrons and positrons in the Galaxy.

\subsection{Revised halo functions for $e^\pm$ in the Galaxy}
\label{sec:halo functions}

We recall that the number density $f(E,r,z)$
of electrons or positrons at the position $(r,z)$ per unit energy $E$ is obtained solving the standard diffusion-loss differential equation
\beq 
\label{eq:diffeq}
- \mathcal{K}_0\, \left(\frac{E}{{\rm GeV}} \right)^\delta\, \nabla^2 f  - \frac{\partial}{\partial E}\Big( b(E,r,z) \, f \Big) = Q(E,r,z),
\eeq
where the first term (which accounts for diffusion) is expressed in terms of the propagation parameters discussed in sec.~\ref{sec:other astro}. The source term $Q$ reads 
\beq 
Q = \left\{ 
\begin{array}{l}
\displaystyle
\frac{1}{2} \left(\frac{\rho}{M_{\rm DM}}\right)^2 \sum_{f} \langle \sigma v\rangle_f \frac{dN_{e^\pm}^f}{dE} \qquad {\rm (annihilation)}\\
\displaystyle
\left(\frac{\rho}{M_{\rm DM}}\right) \sum_{f} \Gamma_f \frac{dN_{e^\pm}^f}{dE} \qquad {\rm (decay)}
\end{array}
\right. .
\label{eq:Q}
\eeq
The function $dN_{e^{\pm}}/dE$ is the electrons or positron spectrum from DM annihilations in a given final state channel $f$.

\begin{figure}[!hp]
\begin{center}
\includegraphics[height= 0.98 \textheight]{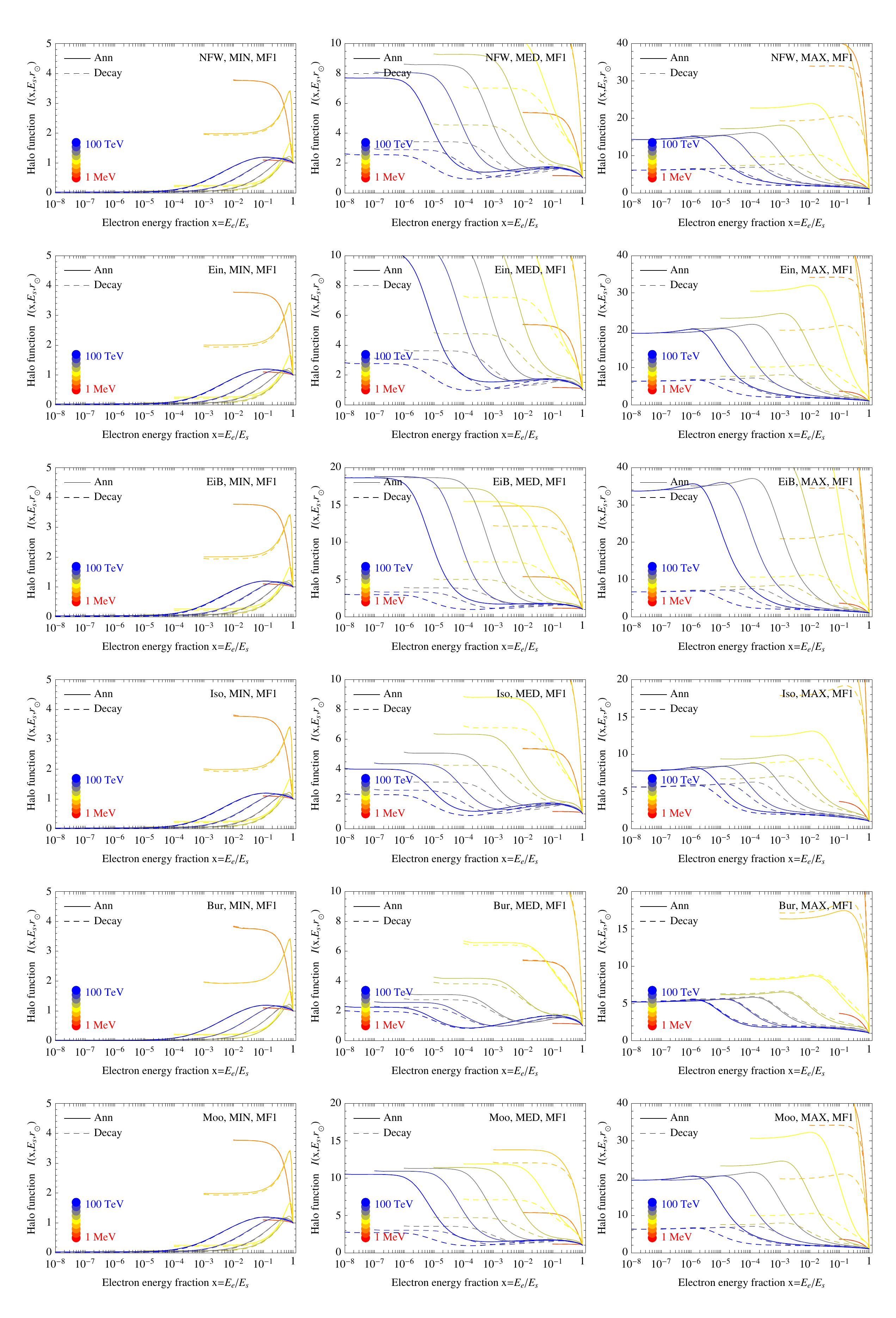}
\vspace{-0.5cm}
\caption{\em \small \label{fig:halofunct2} {\bfseries Generalized halo functions for electrons or positrons}, for several different values of the injection energy $E_s$ (color coded). This figure replaces the analogous one (fig. 6) of~\cite{PPPC4DMID}.}
\end{center}
\end{figure}

\begin{figure}[!t]
\begin{center}
\includegraphics[width=0.48 \textwidth]{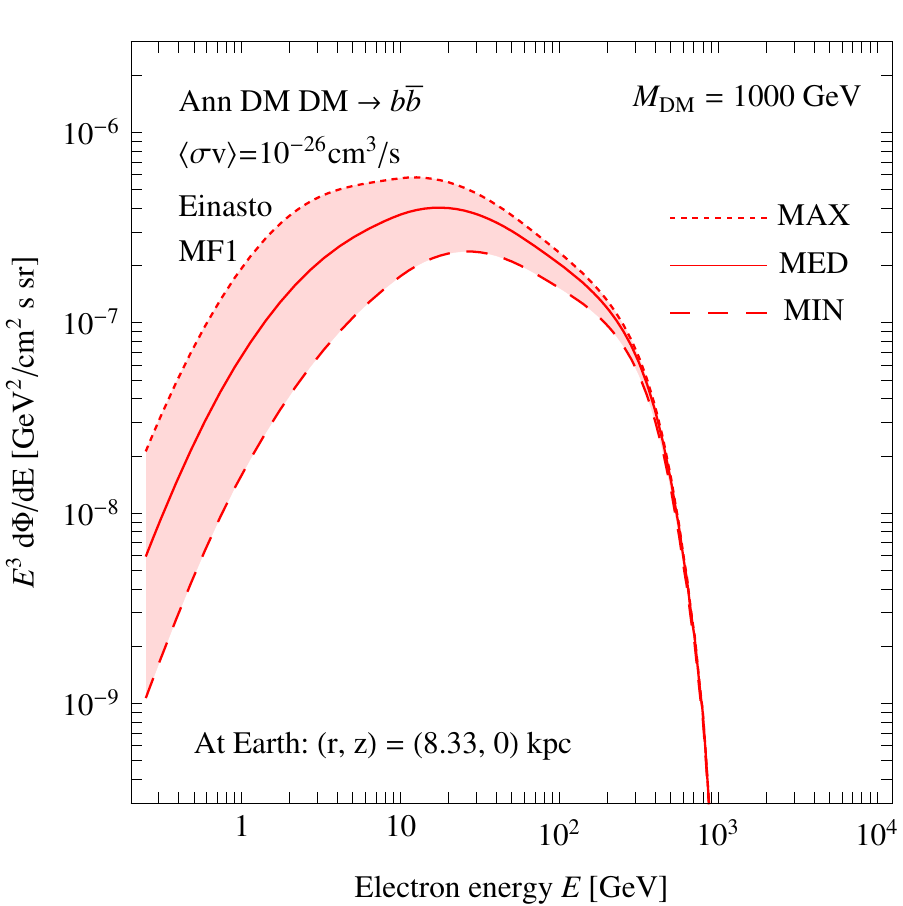} \quad \includegraphics[width=0.48 \textwidth]{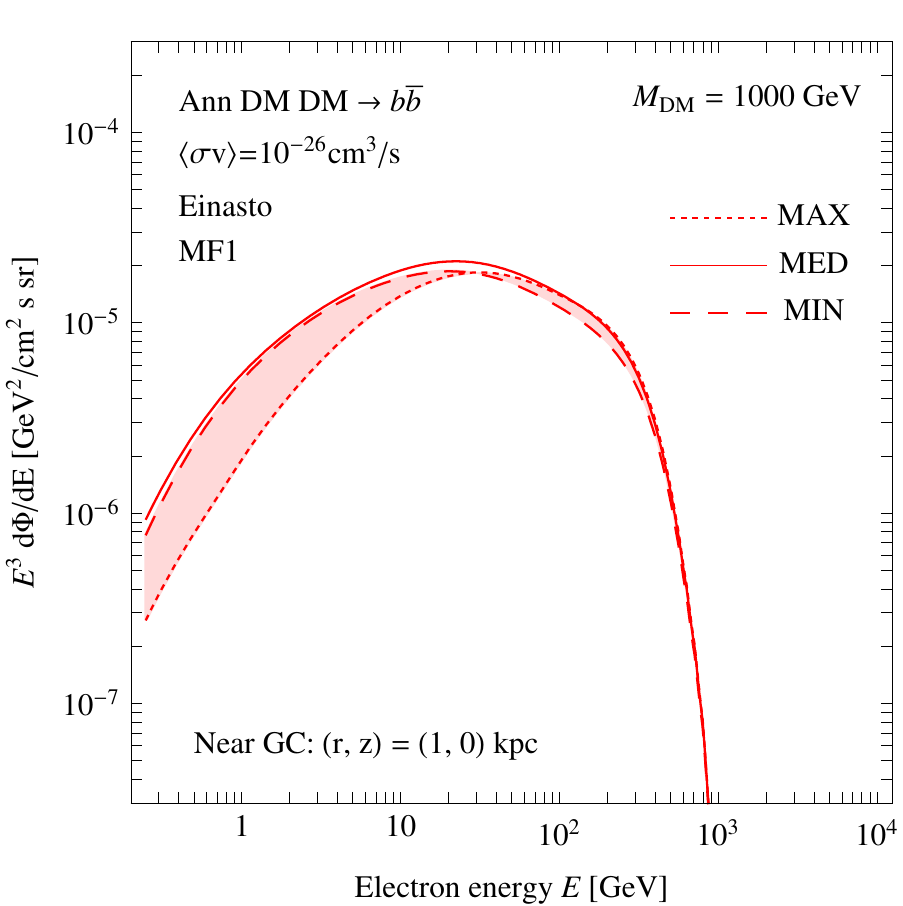}\\
\includegraphics[width=0.48 \textwidth]{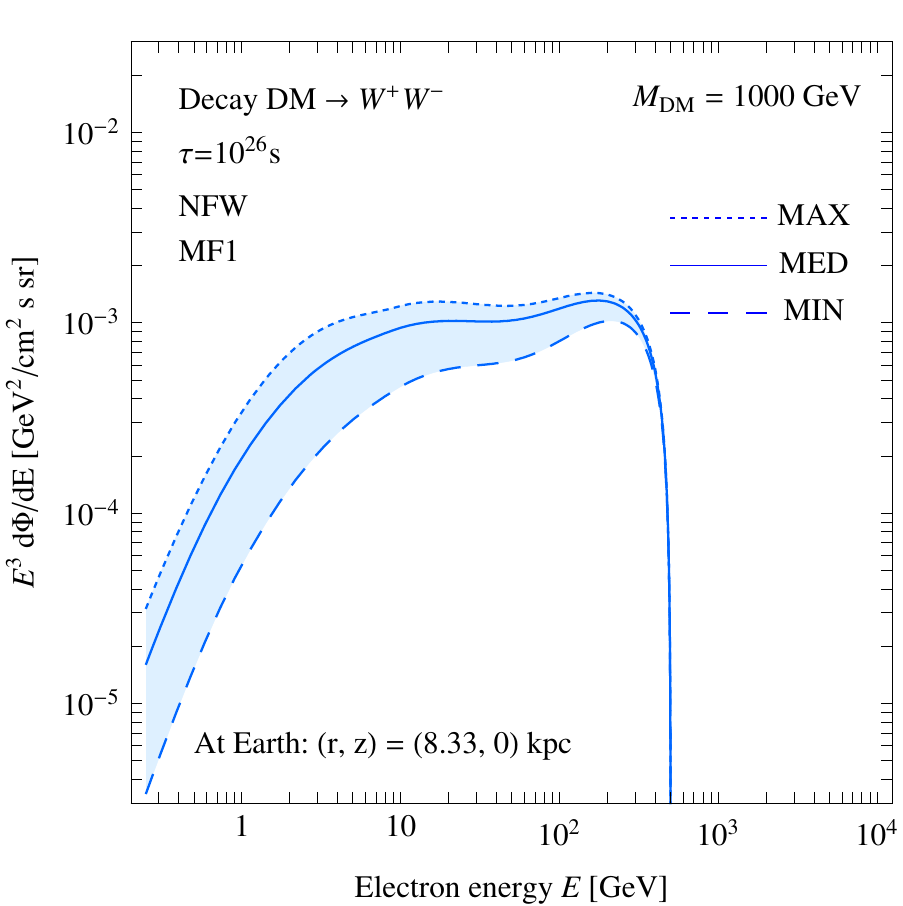} \quad \includegraphics[width=0.48 \textwidth]{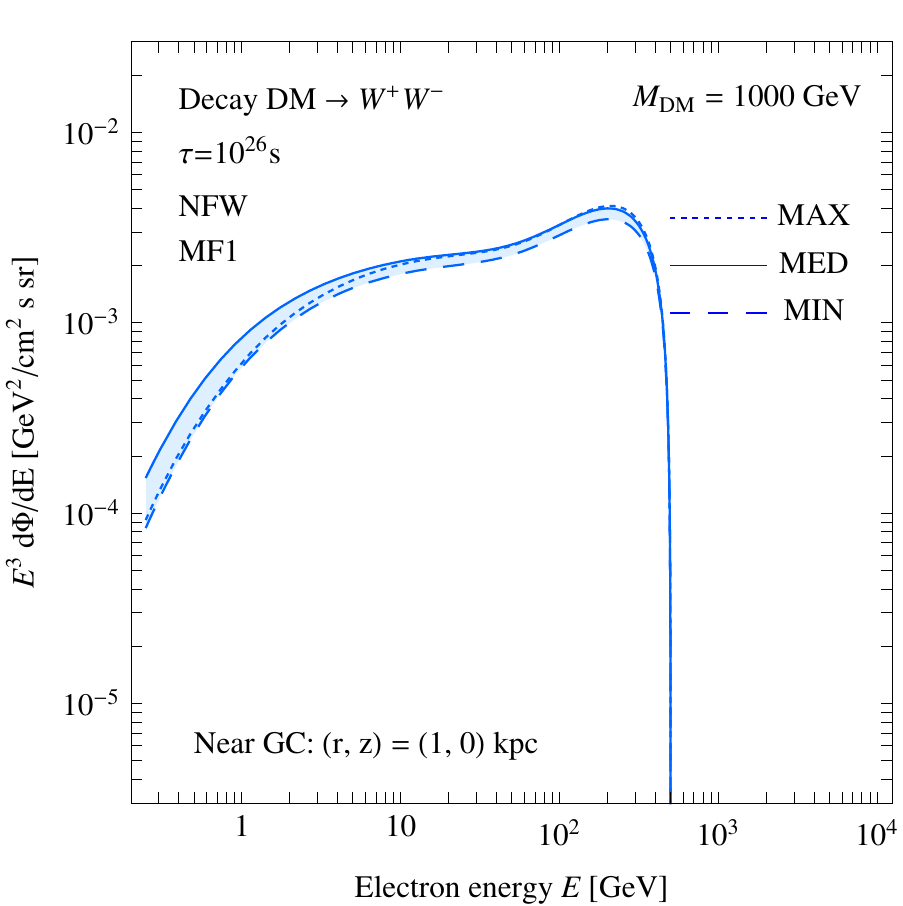}
\caption{\em \small \label{fig:positronspropagated} {\bfseries Fluxes of electrons or positrons, after propagation}, for the case of annihilations (top row) and decay (bottom row), shown at two different locations.}
\end{center}
\end{figure}

The solution for $f$, or rather for the energy spectrum of electrons or positrons $d\Phi_{e^\pm}/dE$, can be cast~\cite{PPPC4DMID} in terms of a convolution of the injection spectrum $dN_{e^{\pm}}/dE$ with the {\it generalized halo functions} $I(E,E_s,r,z)$, which are essentially the Green's functions from a source energy $E_s$ to the energy $E$:
\begin{equation}
\begin{split}
 &\frac{d\Phi_{e^\pm}}{dE}(E,r,z) \equiv \frac{c}{4\pi} f(E,r,z) = \\[5mm]
 & = \frac{c}{4\pi\, b(E,r,z)}  \left\{
 \begin{array}{l}
 \displaystyle\!\! 
 \frac{1}{2}\left(\frac{\rho}{M_{\rm DM}}\right)^2  \sum_f \langle \sigma v \rangle_f  \int_{E}^{M_{\rm DM}} dE_s \frac{dN^f_{e^{\pm}}}{dE}(E_s) \cdot I(E,E_s,r,z) \quad {\rm (annihilation)}\\[5mm]
 \displaystyle\!\! 
 \left(\frac{\rho}{M_{\rm DM}}\right) \sum_f \Gamma_f  \int_{E}^{M_{\rm DM}} dE_s \frac{dN_{e^{\pm}}}{dE}(E_s) \cdot I(E,E_s,r,z) \quad {\rm (decay)}
 \end{array}
 \right. 
 \end{split}
\label{eq:halof}
\end{equation}

The halo functions $I$, which replace those in~\cite{PPPC4DMID}, are available on the~\myurl{www.marcocirelli.net/PPPC4DMID.html}{website}~\cite{website}, in the format \  {\tt ElectronHaloFunctGalaxyAnn[halo,propag,MF][log$_{10}$x,log$_{10}$E$_s$,r,z]} \ (and analogously \ {\tt ElectronHaloFunctGalaxyDec} \ for decay) where $x = E/E_s$.

\medskip

These functions, particularized at the location of the Earth, are plotted in fig.~\ref{fig:halofunct2} for reference. Comparing with the equivalent functions presented in~\cite{PPPC4DMID}, the main difference consists in the evident rise towards small values of the electron energy fraction $x$,\footnote{The functions are defined down to the value of $x$ corresponding to $E = 1$ MeV, to avoid the regime of highly non-relativistic electrons.} which is the direct consequence of the additional, low-energy losses. For small injection energies (warmer colors), the rise occurs `early' while moving towards small $x$, consistently with the fact that the new losses are already relevant. For large injection energies (cold colors) the rise occurs at small $x$ when $E \sim 10$ GeV (the regime at which the new losses set in). For {\sc Min} the rise does not happen for large injection energy, as $e^\pm$ are not efficiently confined on the characteristic scale of the energy losses. At a location closer to the Galactic Center (not plotted), where energy losses are more relevant, the rise is present.

\medskip

In fig.~\ref{fig:positronspropagated} we show the electron spectra, for a few cases. These are in direct correspondence with the left panels of fig.~13 of~\cite{PPPC4DMID}. The differences amount to a factor of a few, up to almost one order of magnitude, especially at low energies (where indeed the new losses are effective). In the right panels of fig.~\ref{fig:positronspropagated} we show the spectra computed in a location closer to the Galactic Center. It is curious to note that, in this case, the fluxes do not follow the intuitive normalization ordering {\sc Min} $\to$ {\sc Med} $\to$ {\sc Max}; in fact, {\sc Max} yields the most suppressed flux. This is just a consequence of the relative importance of the various propagation parameters which is different in the Earth's local neighborhood with respect to that location. Indeed, {\sc Min, Med} and {\sc Max} are determined as the sets that minimize/maximize the fluxes {\em at Earth}.

\subsection{Synchrotron halo functions}
\label{sec:synchr formalism}

In this subsection we want to obtain the {\it generalized halo functions for synchrotron emission} which constitute one of the two main technical outputs of this paper. 
We first review the basics of synchrotron emission and then come to the definition of the functions we need.

\medskip

The synchrotron power (in erg s$^{-1}$ Hz$^{-1}$) emitted in a certain frequency $\nu$ by an isotropic distribution of relativistic electrons with energy $E$ in a uniform magnetic field is

\begin{equation}
\mathcal{P}_{\rm syn}(\nu,E,\alpha)=\sqrt{3}\, \frac{e^3 \, B \sin\alpha}{m_e\, c^2} F(x)
\label{eq:syncreg}
\end{equation}
with $$x=\nu/\nu_c', \hspace{1cm} \nu_c'=\frac12 \nu_c \sin \alpha, \hspace{1cm} \nu_c = \frac{3}{2 \pi}\frac{e}{m_ec} B\gamma^2.$$
Here $B$ is the strength of the magnetic field, $\alpha$ the angle between the line of sight and the magnetic field direction and $\gamma = E/m_e$ the Lorentz factor of the electron or positron.
The synchrotron kernel $F(x)$ is
$$ F(x)=x \int_x^{\infty} K_{5/3}(x^{\prime}) dx^{\prime}$$
where $K_{n}$ is the modified Bessel function of the second kind of order $n$.
In presence of a randomly oriented magnetic field, which is the case of our interest, the synchrotron power has to be averaged over the pitch angle $\alpha$:
\begin{equation}
\mathcal{P}_{\rm syn}(\nu,E)=\frac{1}{2}\int_0^{\pi} d\alpha \, \sin(\alpha) \, \mathcal{P}_{\rm syn}(\nu,E,\alpha)
\label{eq:syncran}
\end{equation}
For relativistic electrons ($\gamma \ge 2$) this corresponds to~\cite{Ghisellini}:
\begin{equation}
\mathcal{P}_{\rm syn}(\nu,E)=2\sqrt{3} \, \frac{e^3\, B}{m_ec^2} y^2 \left[ K_{4/3}(y) K_{1/3}(y)-\frac{3}{5}y\Big( K_{4/3}(y)^2-K_{1/3}(y)^2\Big)\right]
\label{eq:sync1}
\end{equation}
with $y=\nu/\nu_c$.
Integrating this quantity over $\nu$ yields the total power emitted by an electron of energy $E$ in all frequencies, i.e. eq.~(\ref{eq:enlosssyn}).

\medskip

Next, the synchrotron emissivity has to be computed convolving the synchrotron power in eq.~(\ref{eq:sync1}) with the number density of electrons per unit energy $f(E,r,z)$ (in cm$^{-3}$ GeV$^{-1}$) discussed in sec.~\ref{sec:halo functions}  

\begin{equation}
 j_{\rm syn}(\nu,r,z) = 2 \int_{m_e}^{M_{\rm DM}(/2)} dE\ \mathcal{P}_{\rm syn}(\nu,E)\,  f(E,r,z)
\label{eq:sync2}
\end{equation}
where the minimal and maximal energies of the emitting electrons are determined by the electron mass and the mass of the DM particle. The `/2' notation applies to the decay case. The overall factor 2 takes into account that, besides the electrons, an equal population of positrons radiates.

\medskip

Finally, the observable in which we are interested is the intensity $\mathfrak{I}$ of the synchrotron emission (in erg cm$^{-2}$ s$^{-1}$ Hz$^{-1}$ sr$^{-1}$) from a certain direction of observation. This is obtained by integrating the emissivity of eq.~(\ref{eq:sync2}) along the line-of-sight. Schematically:

\begin{equation}
\mathfrak{I}(\nu,b,\ell) = \int_{\rm l.o.s.}ds \, \frac{ j_{\rm syn}(\nu,r,z)}{4\pi} 
\label{eq:sync3}
\end{equation}
where it is intended that a point in $(r,z)$ is identified by the the parameter $s$ along the line of observation individuated by the galactic latitude $b$ and longitude $\ell$: $r(s,\ell,b), z(s,\ell,b)$~\footnote{We remind the explicit relations $r(s,\ell,b) = s \sin b, z(s,\ell,b) = \sqrt{r_\odot^2+s^2-2\, r_\odot \cos b \cos \ell}$.}.
\medskip

\begin{figure}[!hp]
\begin{center}
\includegraphics[height= 0.95 \textheight]{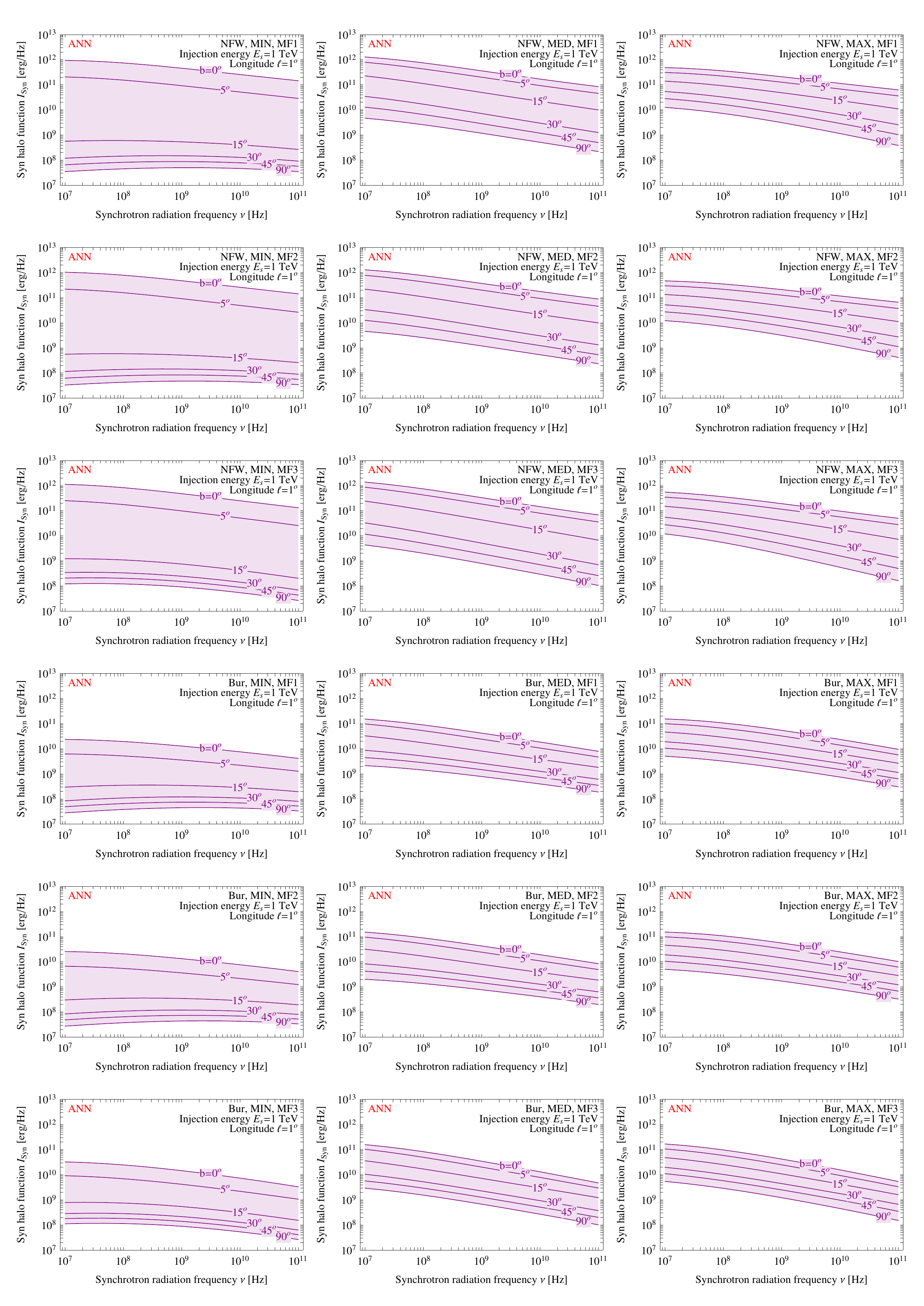}
\vspace{-0.7cm}
\caption{\em \small \label{fig:IsynAnn} {\bfseries Generalized synchrotron halo functions}, for the DM annihilation case. The upper 9 panels correspond to an NFW profile, the lower 9 to Burkert; the columns correspond to a fixed propagation model, the rows to a fixed magnetic field model.}
\end{center}
\end{figure}

\begin{figure}[!hp]
\begin{center}
\includegraphics[height= 0.95 \textheight]{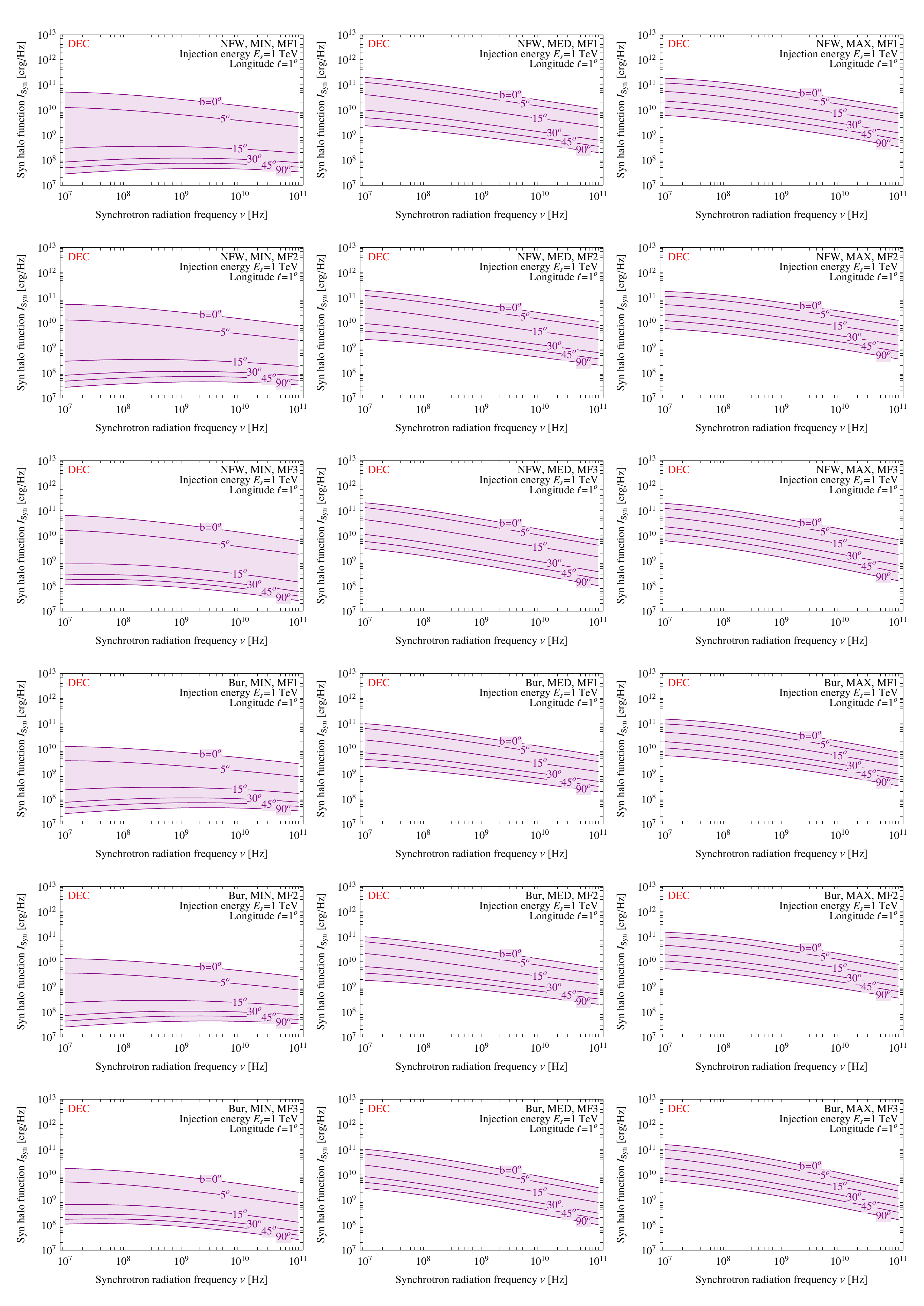}
\vspace{-0.6cm}
\caption{\em \small \label{fig:IsynDec} {\bfseries Generalized synchrotron halo functions}, for the DM decay case.}
\end{center}
\end{figure}

Recollecting eq.~(\ref{eq:sync3}) and eq.~(\ref{eq:halof}), the synchrotron intensity $\mathfrak{I}$ at a given frequency $\nu$ and for given galactic coordinates $(b,\ell)$ can be cast as:

\begin{equation}
 \mathfrak{I}(\nu,\ell,b)=\frac{r_\odot}{4\pi} \left\{
 \begin{array}{l}
 \displaystyle\!\!
 \frac{1}{2}\left(\frac{\rho_{\odot}}{M_{\rm DM}}\right)^2  \int_{m_e}^{M_{\rm DM}} dE_s  \sum_f \langle \sigma v \rangle_f \frac{dN^f_{e^{\pm}}}{dE}(E_s)\ I_{\rm syn}(E_s,\nu,\ell,b) \quad {\rm (annihilation)}\\
\displaystyle\!\!
 \left(\frac{\rho_{\odot}}{M_{\rm DM}}\right) \int_{m_e}^{M_{\rm DM}/2} dE_s  \sum_f \Gamma_f \frac{dN^f_{e^{\pm}}}{dE}(E_s)\ I_{\rm syn}(E_s,\nu,\ell,b) \quad {\rm (decay)}
 \end{array}
 \right.
\label{eq:synchintensity}
\end{equation}
with the {\it generalized synchrotron halo function} $I_{\rm syn}(\nu,E_s,\ell,b)$ defined as
\begin{equation}
I_{\rm syn}(E_s,\nu,\ell,b)= \int_{\rm l.o.s.}\frac{ds}{r_\odot} \left(\frac{\rho(r,z)}{\rho_\odot}\right)^\eta \  2 \int_{m_e}^{E_s} dE  \frac{\mathcal{P}_{\rm syn}(\nu,E)}{b(E,r,z)} \ I(E,E_s,r,z),
\label{eq:synchalof}
\end{equation}
where $\eta = 1,2$ for the decay or annihilation cases respectively and again implicitly $r(s,\ell,b)$, $z(s,\ell,b)$. The units of $I_{\rm syn}$ are erg/Hz.

\medskip

The synchrotron halo functions $I_{\rm syn}$ are available on the~\myurl{www.marcocirelli.net/PPPC4DMID.html}{website}~\cite{website}, in the format \  {\tt ISynAnnI[halo,propag,MF][log$_{10}$E$_{s}$,log$_{10}$$\nu$,$\ell$,$b$]} \ (and analogously \ {\tt ISynDecI} \ for decay). They are also plotted, for reference, in fig.~\ref{fig:IsynAnn} for the annihilation case and in fig.~\ref{fig:IsynDec} for the decay case.

\bigskip

\begin{figure}[t]
\begin{center}
\includegraphics[width= 0.7 \textwidth]{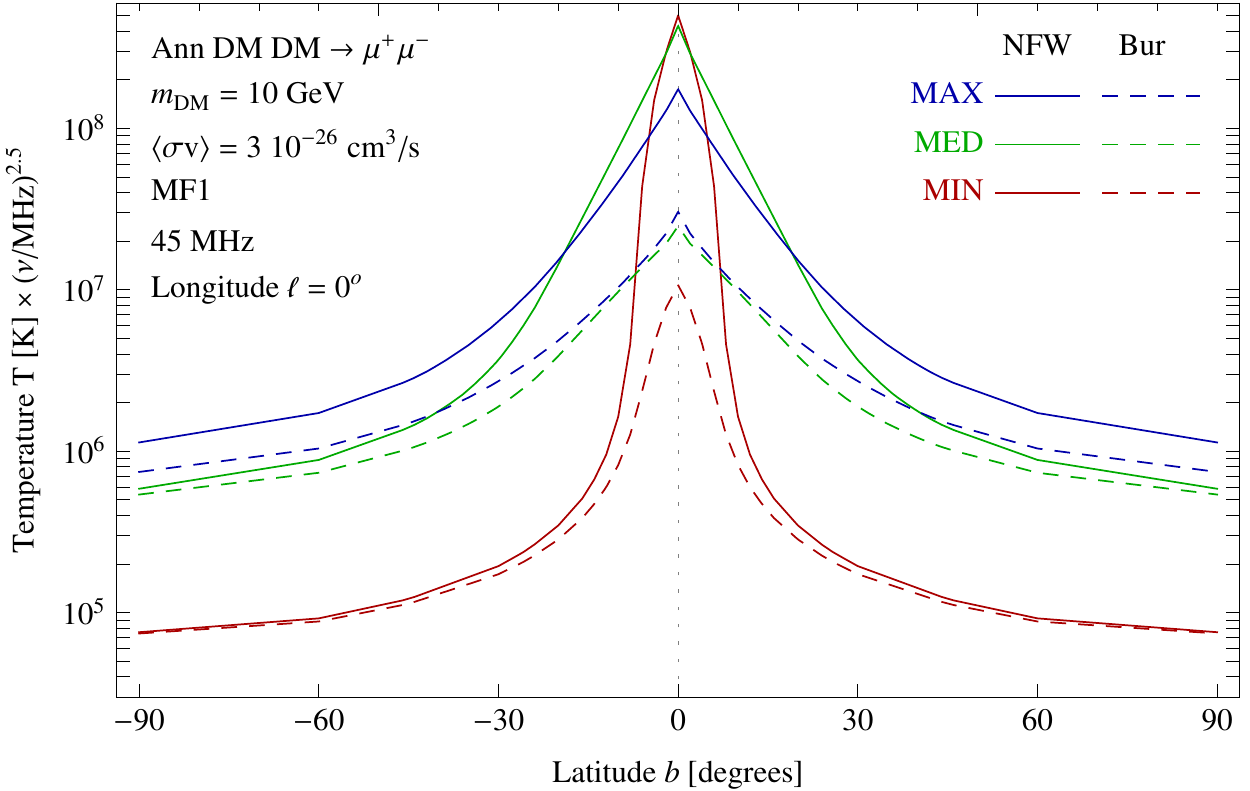}
\caption{\em \small \label{fig:synchflux} {\bfseries Synchrotron signal} (temperature)  at 45 MHz, plotted against the galactic latitude, for several choices of DM profile and propagation scheme.}
\end{center}
\end{figure}

The last step needed in order to make contact with actual radio surveys consists in expressing the synchrotron signal in terms of brightness temperature $T(\nu)$ (in K) which is defined as:
\begin{equation}
 T(\nu)=\frac{c^2 \, \mathfrak{I}(\nu)}{2\, \nu^2 \, k_B}
\label{eq:syncT}
\end{equation}
with $k_B$ the Boltzmann constant. In fig.~\ref{fig:synchflux} we plot such quantity for a few different choices of profiles and propagation parameters.
Although a comparison with previous results (e.g. in~\cite{Fornengo:2011iq} and~\cite{Mambrini:2012ue}) is not possible in full details, we have checked that, removing our additional refinements, we recover those previous results in most cases.~\footnote{We cannot however fully reproduce the dependence on the choice of profile in~\cite{Mambrini:2012ue}: we find that the synchrotron signal is independent on the choice of profile at large latitudes (as we expect from the self-similarity of such profiles at large radii) while their plots show a sizable residual difference.}

\medskip

Before moving on, we would like to point out that our tools can be adapted for usage in a more general way. Notably, if a user is interested in the synchrotron signatures from a custom galactic magnetic field configuration, she can employ our electron and positron galactic density and fold it with the desired MF. This is formally not self-consistent (one would be computing the synchrotron energy losses with one MF configuration but the synchrotron emission with a different one), but it can be acceptable for practical purposes in the conditions in which the dominant energy losses are due to other processes like ICS and bremsstrahlung (corresponding to high energies or to regions where the magnetic field is not too large, see the discussion in Sec.~\ref{sec:energy loss}). Technically, one needs to compute the quantity $f(E,r,z)$ as presented in eq.~(\ref{eq:halof}) using the electron and positron halo functions provided in Sec.~\ref{sec:halo functions}. The quantity $f$ can then just be plugged in eq.~(\ref{eq:sync2}) and then eq.~(\ref{eq:sync3}) to compute the synchrotron emission $\mathfrak{I}$. The custom configuration of the MF enters in determining the corresponding synchrotron power with eq.~(\ref{eq:sync1}).


\subsection{Bremsstrahlung halo functions}
\label{sec:brem formalism}

In this subsection, in turn, we want to obtain the {\it generalized halo functions for bremsstrahlung emission}, the other main technical output of this paper. 
The computation follows quite closely the one for synchrotron in the previous subsection, using also the formalism for bremsstrahlung spelled out in sec.~\ref{sec:energy loss}. We summarize here the main ingredients for completeness.

\medskip

\begin{figure}[!hp]
\begin{center}
\includegraphics[height= 0.95 \textheight]{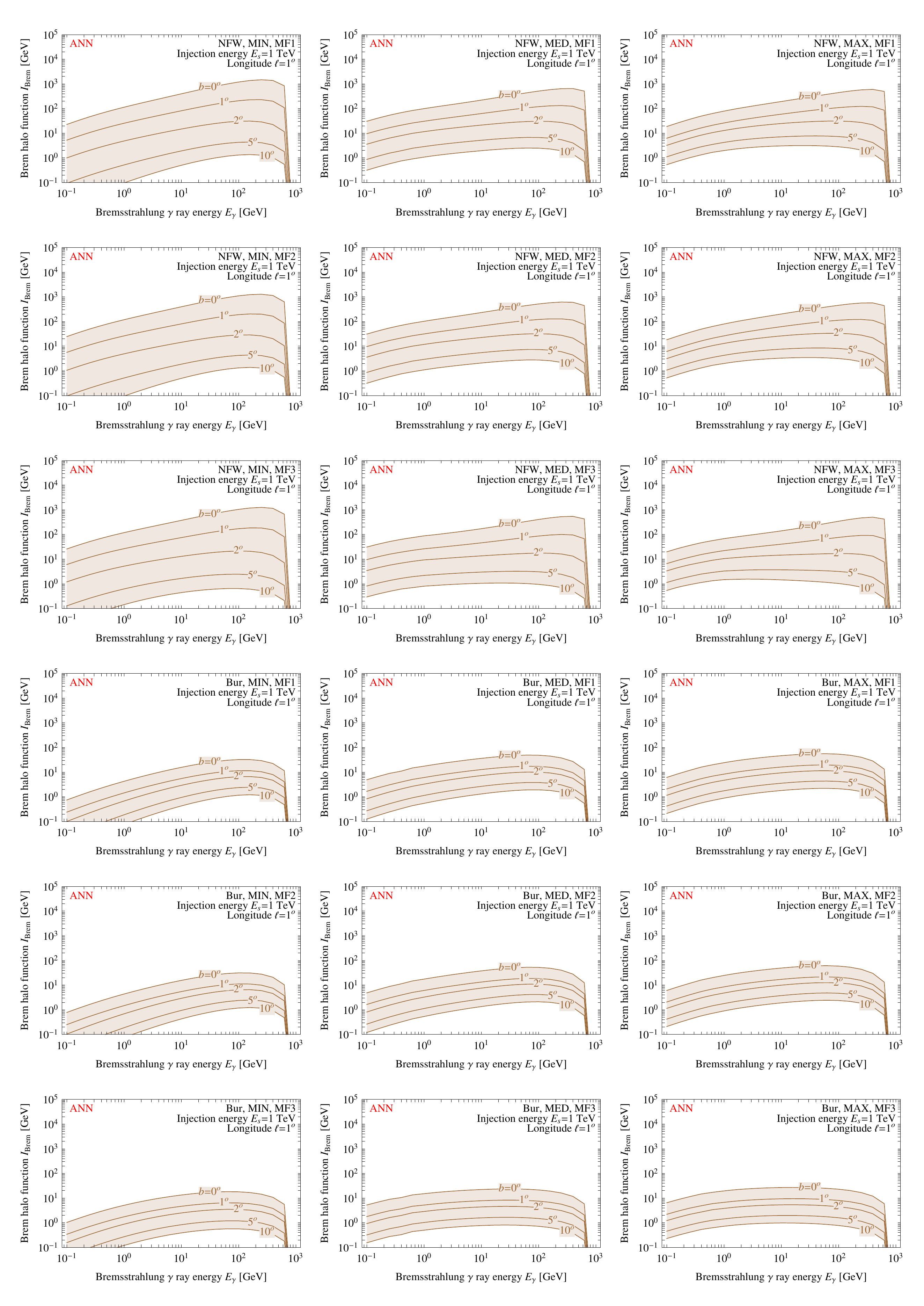}
\vspace{-0.7cm}
\caption{\em \small \label{fig:IbremAnn} {\bfseries Generalized bremsstrahlung halo functions}, for the DM annihilation case. Analogously to fig.~\ref{fig:IsynAnn}, the upper 9 panels correspond to an NFW profile, the lower 9 to Burkert; the columns correspond to a fixed propagation model, the rows to a fixed magnetic field model.}
\end{center}
\end{figure}

\begin{figure}[!hp]
\begin{center}
\includegraphics[height= 0.95 \textheight]{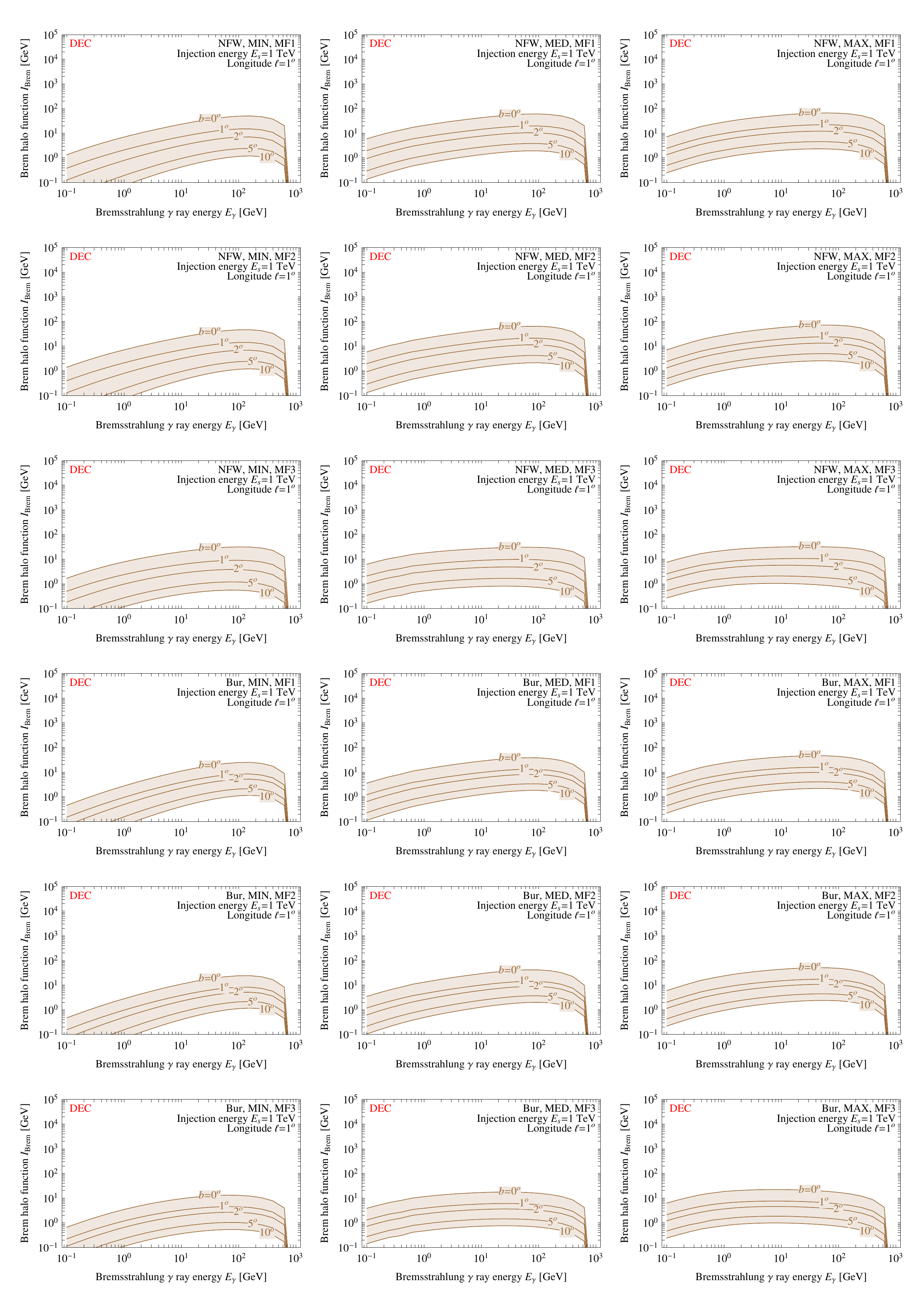}
\vspace{-0.7cm}
\caption{\em \small \label{fig:IbremDec} {\bfseries Generalized bremsstrahlung halo functions}, for the DM decay case.}
\end{center}
\end{figure}

\begin{figure}[!t]
\begin{center}
\includegraphics[width= 0.48 \textwidth]{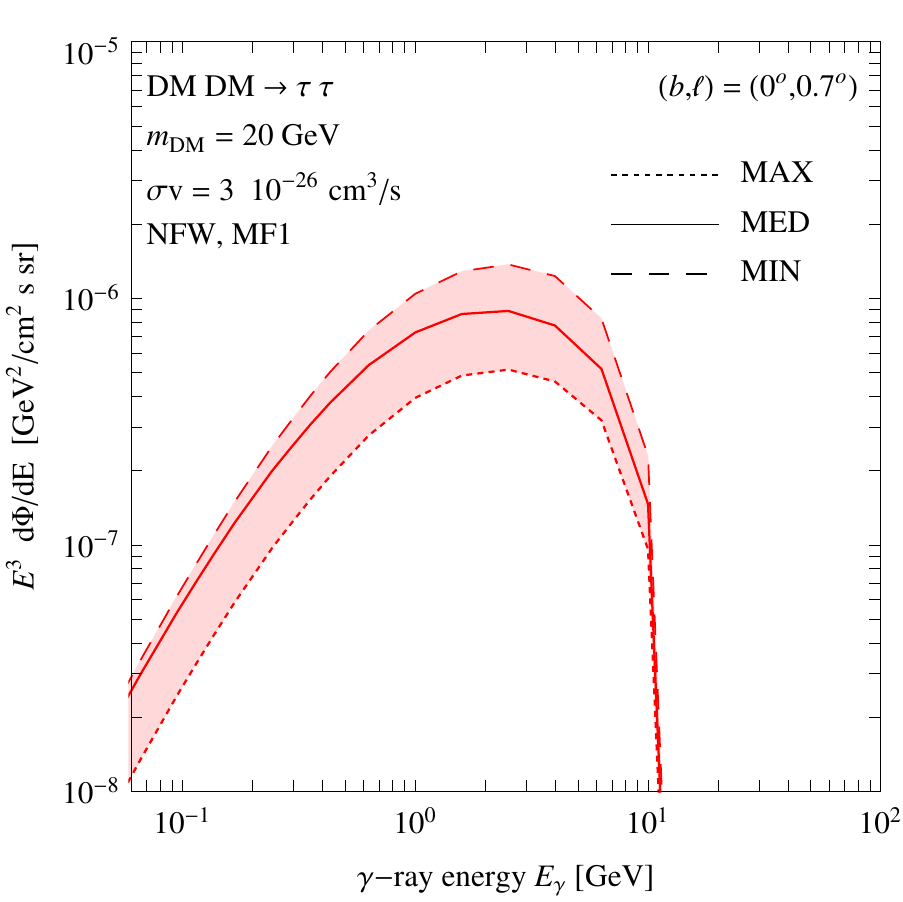} \quad \includegraphics[width= 0.48 \textwidth]{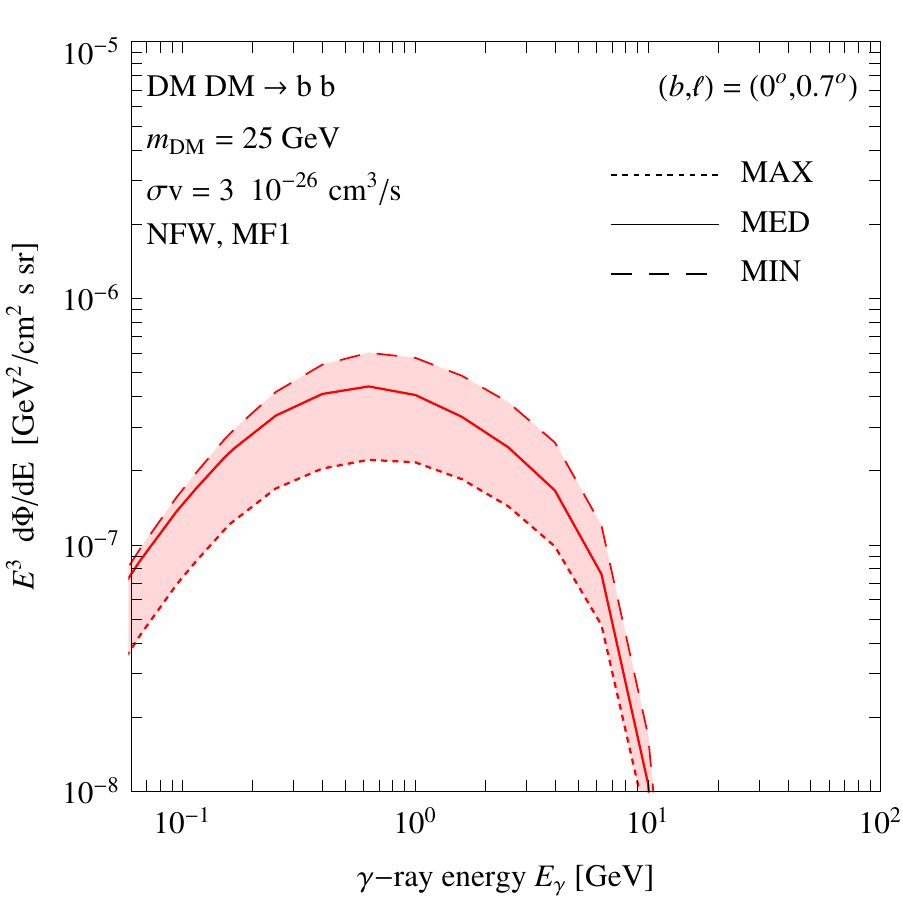}\\[2mm]
\includegraphics[width= 0.48 \textwidth]{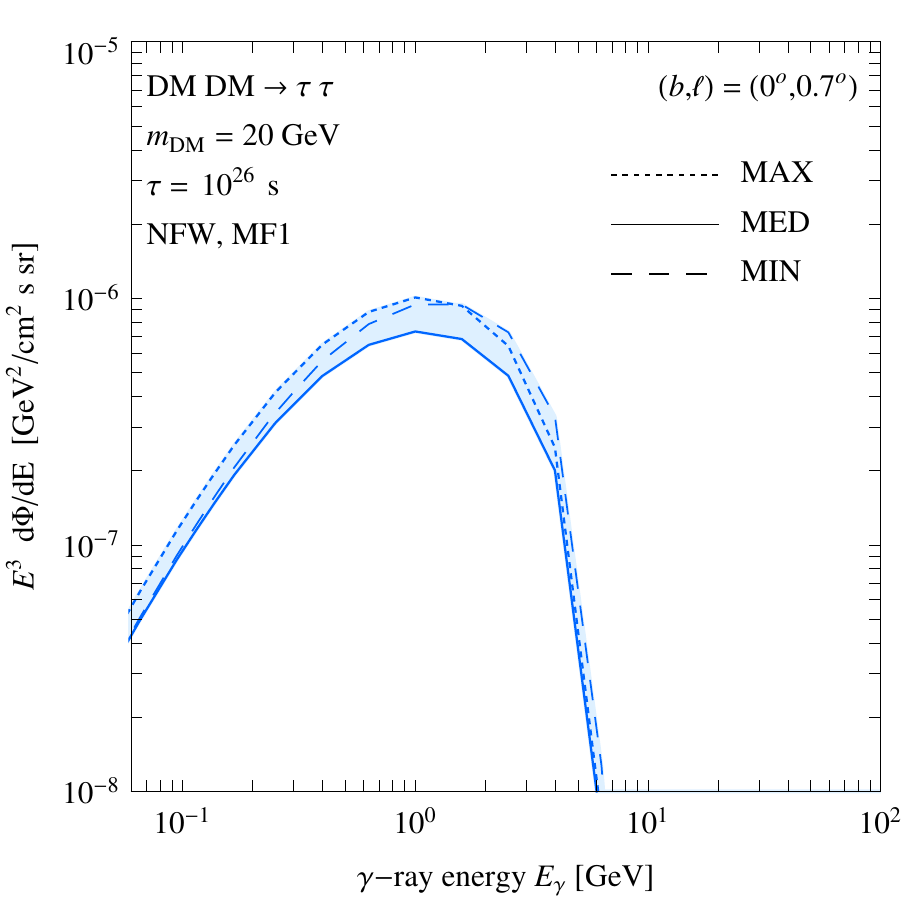} \quad \includegraphics[width= 0.48 \textwidth]{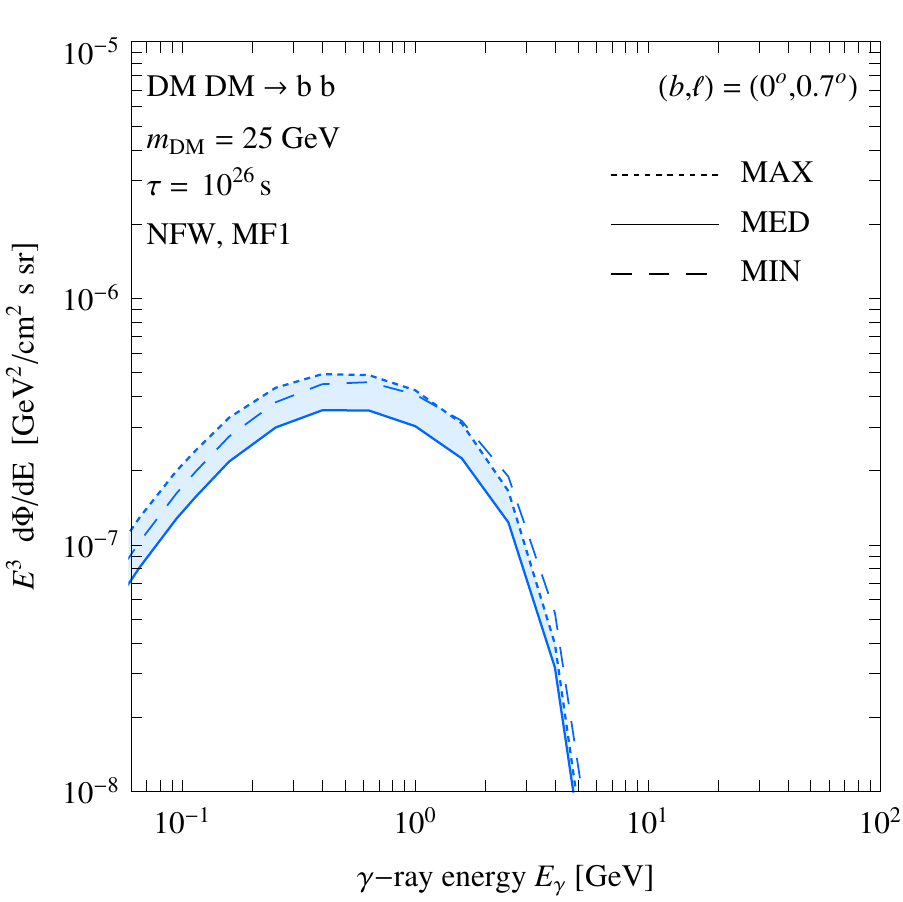}
\caption{\em \small \label{fig:bremflux} {\bfseries Bremsstrahlung $\gamma$-ray fluxes} for the case of annihilations (top row) and decay (bottom row), shown for two different channels.}
\end{center}
\end{figure}

In close analogy with eq.~(\ref{eq:synchintensity}), the bremsstrahlung differential flux (in GeV$^{-1} {\rm cm}^{-2} {\rm s}^{-1} {\rm sr}^{-1}$) reads: 
\beq
\frac{d\Phi_{{\rm brem}\gamma}}{dE_\gamma\, d\Omega} = \frac1{E_\gamma^2}\frac{r_\odot}{4\pi} 
\left\{\begin{array}{l}
\displaystyle \frac12 \left(\frac{\rho_\odot}{M_{\rm DM}}\right)^2 \int_{m_e}^{M_{\rm DM}} \hspace{-0.45cm} dE_{\rm s}  \sum_f \langle \sigma v \rangle_f \frac{dN^f_{e^\pm}}{dE}(E_{\rm s})  \, I_{\rm brem}(E_{\rm s},E_\gamma,\ell,b) \quad {\rm (annihilation)} \\[5mm]
\displaystyle \frac{\rho_\odot}{M_{\rm DM}} \int_{m_e}^{M_{\rm DM}/2}\hspace{-0.45cm} dE_{\rm s}  \sum_f \Gamma_f  \frac{dN^f_{e^\pm}}{dE}(E_{\rm s})  \, I_{\rm brem}(E_{\rm s},E_\gamma,\ell,b) \qquad \,\,\,\,\,{\rm (decay)}
\end{array}\right.
\label{eq:summarbrem}
\eeq
where now (in analogy with eq.~(\ref{eq:synchalof})) the {\it generalized halo function for bremsstrahlung} $I_{\rm brem}$ $(E_{\rm s},E_\gamma,\ell,b)$,  which has units of GeV, is defined as 
\beq\label{Halobrem}
I_{\rm brem}(E_{\rm s},E_\gamma,\ell,b)= 2\, E_\gamma \int_{\rm l.o.s.} \frac{ds}{r_\odot} \left( \frac{\rho(r,z)}{\rho_\odot} \right)^\eta \hspace{-0.2cm} \int_{m_e}^{E_{\rm s}}\hspace{-0.3cm}dE\,  \frac{{\mathcal P_{\rm brem}(E_\gamma,E,r(s,\theta))}}{b(E,r(s,\theta))}\,  I(E,E_{\rm s},r,z).
\eeq
The bremsstrahlung power consists in
\begin{equation}
{\mathcal P}_{\rm brem} (E_\gamma, E, \vec{x}) = c \, E_\gamma \sum_i n_i(\vec{x}) \frac{d \sigma_i(E_\gamma,E)}{dE_\gamma}
\end{equation}
where $n_i$ are the number densities of the gas species and the bremsstrahlung cross section was given in eq.~(\ref{eq:sigma}).

\bigskip

The bremsstrahlung halo functions $I_{\rm brem}$ are again available on the~\myurl{www.marcocirelli.net/PPPC4DMID.html}{website}~\cite{website}, in the format \  {\tt IBremAnnI[halo,propag,MF][log$_{10}$E$_{s}$,log$_{10}$E$_\gamma$,$\ell$,$b$]} \ (and analogously \ {\tt IBremDecI} \ for decay). 
They are plotted in fig.~\ref{fig:IbremAnn} and~\ref{fig:IbremDec} (annihilation and decay cases), for reference. 

\bigskip

In fig.~\ref{fig:bremflux} we plot the resulting bremsstrahlung line-of-sight $\gamma$-ray fluxes, for a few cases. The agreement with previous calculations (notably~\cite{Cirelli:2013mqa}) has been verified. We also cross checked with fully numerical computations done using {\tt GammaSky}~\cite{DiBernardo:2012zu,Evoli:2012ha}. While the spectral shape is in very good agreement, we find a difference in overall normalization of the fluxes along lines of sight passing close to the Galactic Center. This is due to the fact that {\tt GammaSky}, like {\tt GalProp}, corrects the bremsstrahlung emissivities by adjusting the normalization of the gas densities in each galactocentric ring, in particular close to the Galactic Center (see~\cite{FermiLAT:2012aa}, and~\cite{Cirelli:2013mqa} for a short discussion). We decide to instead use consistently the same maps for energy losses and bremsstrahlung emission.

\section{Conclusions}
\label{sec:conclusions}

In the quest for the discovery of Dark Matter, it is important to exploit all possible strategies. In Indirect searches, in addition, it is important to be able to exploit the multi-messenger and multi-wavelength nature of the possible signals.  
We have here focussed on the secondary radiations from electrons and positrons and presented several upgraded and new results. The upgradings concern: i) an improved energy loss function (Sec.~\ref{sec:energy loss}) which fully includes low energy losses (Coulomb, ionization and bremsstrahlung) and ii) the revised halo functions for electrons and positrons (Sec.~\ref{sec:halo functions}). The new results consist in: iii) the synchrotron halo functions (Sec.~\ref{sec:synchr formalism}); iv) the bremsstrahlung halo functions (Sec.~\ref{sec:brem formalism}). All the results are provided in numerical form on the \myurl{www.marcocirelli.net/PPPC4DMID.html}{{\sc Pppc4dmid} website}~\cite{website}.

These state-of-the-art tools allow to compute the secondary radiation signal (synchrotron, bremsstrahlung and Inverse Compton) from any arbitrary DM weak-scale model and will be precious and hopefully instrumental in the current era of precision DM indirect searches.

\bigskip

{\footnotesize
\paragraph{Acknowledgements}

We thank C\'eline B\oe hm, Alessandro Cuoco and Gabrijela Zaharijas for several valuable inputs, and we are in particular grateful to Paolo Panci for many useful discussions. We acknowledge the hospitality of the Institut d'Astrophysique de Paris, where part of this work was done.

\noindent Funding and research infrastructure acknowledgements: 
\begin{itemize}
\item[$\ast$] European Research Council ({\sc Erc}) under the EU Seventh Framework Programme (FP7/2007-2013)/{\sc Erc} Starting Grant (agreement n.\ 278234 --- `{\sc NewDark}' project),
\item[$\ast$] EU ITN network {\sc Unilhc} [work of M.C.], 
\item[$\ast$] French national research agency {\sc Anr} under contract {\sc Anr} 2010 {\sc Blanc} 041301.
 \end{itemize}
}

\medskip

\footnotesize
\begin{multicols}{2}
  
\end{multicols}


\begin{thebibliography}{nn}
  
\bibitem{radioGC}

  R.~Aloisio, P.~Blasi and A.~V.~Olinto,
  JCAP {\bf 0405} (2004) 007
  [\hhref{astro-ph/0402588}].
  G.~Bertone, E.~Nezri, J.~Orloff and J.~Silk,
  Phys.\ Rev.\ D {\bf 70} (2004) 063503
  [\hhref{astro-ph/0403322}].
M.~Regis and P.~Ullio,
  Phys.\ Rev.\ D {\bf 78} (2008) 043505
  [arXiv:\hhref{0802.0234} [hep-ph]].
G.~Bertone, M.~Cirelli, A.~Strumia and M.~Taoso,
  JCAP {\bf 0903} (2009) 009
  [arXiv:\hhref{0811.3744} [astro-ph]]. 
L.~Bergstrom, G.~Bertone, T.~Bringmann, J.~Edsjo and M.~Taoso,
  Phys.\ Rev.\ D {\bf 79} (2009) 081303
  [arXiv:\hhref{0812.3895} [astro-ph]].
R.~M.~Crocker, N.~F.~Bell, C.~Balazs and D.~I.~Jones,
  Phys.\ Rev.\ D {\bf 81} (2010) 063516
  [arXiv:\hhref{1002.0229} [hep-ph]]. 
  C.~Boehm, J.~Silk and T.~Ensslin,
  arXiv:\hhref{1008.5175} [astro-ph.GA].
  I.~Cholis, D.~Hooper and T.~Linden,
  arXiv:\hhref{1408.6224} [astro-ph.HE].
  
  \bibitem{Blasi:2002ct}
  P.~Blasi, A.~V.~Olinto and C.~Tyler,
  Astropart.\ Phys.\  {\bf 18} (2003) 649
  [\hhref{astro-ph/0202049}].
  
\bibitem{Fornengo:2011iq}
  N.~Fornengo, R.~A.~Lineros, M.~Regis and M.~Taoso,
  JCAP {\bf 1201} (2012) 005
  [arXiv:\hhref{1110.4337} [astro-ph.GA]].
  
\bibitem{Borriello:2008gy}
  E.~Borriello, A.~Cuoco and G.~Miele,
  Phys.\ Rev.\ D {\bf 79} (2009) 023518
  [arXiv:\hhref{0809.2990} [astro-ph]].

\bibitem{Borriello:2008dt}
  E.~Borriello, A.~Cuoco and G.~Miele,
  Nucl.\ Phys.\ Proc.\ Suppl.\  {\bf 190} (2009) 185
  [arXiv:\hhref{0812.2932} [astro-ph]].

\bibitem{Delahaye:2011jn}
  T.~Delahaye, C.~Boehm and J.~Silk,
  Mon.\ Not.\ Roy.\ Astron.\ Soc.\ Lett.\  {\bf 422} (2012) L16
  [arXiv:\hhref{1105.4689} [astro-ph.GA]].

\bibitem{Linden:2011au}
  T.~Linden, D.~Hooper and F.~Yusef-Zadeh,
  Astrophys.\ J.\  {\bf 741} (2011) 95
  [arXiv:\hhref{1106.5493} [astro-ph.HE]].
  
\bibitem{Mambrini:2012ue}
  Y.~Mambrini, M.~H.~G.~Tytgat, G.~Zaharijas and B.~Zaldivar,
  JCAP {\bf 1211} (2012) 038
  [arXiv:\hhref{1206.2352} [hep-ph]].

\bibitem{Fornengo:2011cn}
  N.~Fornengo, R.~Lineros, M.~Regis and M.~Taoso,
  Phys.\ Rev.\ Lett.\  {\bf 107} (2011) 271302
  [arXiv:\hhref{1108.0569} [hep-ph]].

\bibitem{Fornengo:2011xk}
  N.~Fornengo, R.~Lineros, M.~Regis and M.~Taoso,
  JCAP {\bf 1203} (2012) 033
  [arXiv:\hhref{1112.4517} [astro-ph.CO]].

\bibitem{Hooper:2012jc}
  D.~Hooper, A.~V.~Belikov, T.~E.~Jeltema, T.~Linden, S.~Profumo and T.~R.~Slatyer,
  Phys.\ Rev.\ D {\bf 86} (2012) 103003
  [arXiv:\hhref{1203.3547} [astro-ph.CO]].

\bibitem{Carlson:2012qc}
  E.~Carlson, D.~Hooper, T.~Linden and S.~Profumo,
  JCAP {\bf 1307} (2013) 026
  [arXiv:\hhref{1212.5747} [astro-ph.CO]].

\bibitem{Cirelli:2013mqa}
M.~Cirelli, P.~D.~Serpico and G.~Zaharijas,
  JCAP {\bf 1311} (2013) 035
  [arXiv:\hhref{1307.7152}].

\bibitem{Abazajian:2014fta}
  K.~N.~Abazajian, N.~Canac, S.~Horiuchi and M.~Kaplinghat,
  Phys.\ Rev.\ D {\bf 90} (2014) 2,  023526
  [arXiv:\hhref{1402.4090} [astro-ph.HE]].
  
  \bibitem{Daylan:2014rsa}
  T.~Daylan, D.~P.~Finkbeiner, D.~Hooper, T.~Linden, S.~K.~N.~Portillo, N.~L.~Rodd and T.~R.~Slatyer,
  arXiv:\hhref{1402.6703} [astro-ph.HE].
  
  \bibitem{Lacroix:2014eea}
  T.~Lacroix, C.~Boehm and J.~Silk,
  Phys.\ Rev.\ D {\bf 90} (2014) 4,  043508
  [arXiv:\hhref{1403.1987}].
  
 \bibitem{Cirelli:2014lwa}
  M.~Cirelli, D.~Gaggero, G.~Giesen, M.~Taoso and A.~Urbano,
  arXiv:\hhref{1407.2173} [hep-ph].
   
   \bibitem{Abazajian:2014hsa}
  K.~N.~Abazajian, N.~Canac, S.~Horiuchi, M.~Kaplinghat and A.~Kwa,
  arXiv:\hhref{1410.6168} [astro-ph.HE].
  
  \bibitem{Eichler:2005aa}
  D.~Eichler and I.~Maor,
  \hhref{astro-ph/0501096}.
  
  \bibitem{Cholis:2008wq}
  I.~Cholis, G.~Dobler, D.~P.~Finkbeiner, L.~Goodenough and N.~Weiner,
  Phys.\ Rev.\ D {\bf 80} (2009) 123518
  [arXiv:\hhref{0811.3641} [astro-ph]].
  
  \bibitem{Zhang:2008tb}
  J.~Zhang, X.~J.~Bi, J.~Liu, S.~M.~Liu, P.~F.~Yin, Q.~Yuan and S.~H.~Zhu,
  Phys.\ Rev.\ D {\bf 80} (2009) 023007
  [arXiv:\hhref{0812.0522} [astro-ph]].
  
  \bibitem{Borriello:2009fa}
  E.~Borriello, A.~Cuoco and G.~Miele,
  Astrophys.\ J.\  {\bf 699} (2009) L59
  [arXiv:\hhref{0903.1852} [astro-ph.GA]].
  
  \bibitem{Barger:2009yt}
  V.~Barger, Y.~Gao, W.~Y.~Keung, D.~Marfatia and G.~Shaughnessy,
  Phys.\ Lett.\ B {\bf 678} (2009) 283
  [arXiv:\hhref{0904.2001} [hep-ph]].
  
  \bibitem{Cirelli:2009vg}
  M.~Cirelli and P.~Panci,
  Nucl.\ Phys.\ B {\bf 821} (2009) 399
  [arXiv:\hhref{0904.3830} [astro-ph.CO]].
  
\bibitem{PPPC4DMID}  
M.~Cirelli, G.~Corcella, A.~Hektor, G.~Hutsi, M.~Kadastik, P.~Panci, M.~Raidal and F.~Sala {\it et al.},
  JCAP {\bf 1103} (2011) 051
   [Erratum-ibid.\  {\bf 1210} (2012) E01]
  [arXiv:\hhref{1012.4515} [hep-ph]].

\bibitem{PPPC4DMnu}P.~Baratella, M.~Cirelli, A.~Hektor, J.~Pata, M.~Piibeleht and A.~Strumia,
  JCAP {\bf 1403} (2014) 053
  [arXiv:\hhref{1312.6408} [hep-ph]].

\bibitem{future_synchr}
M.~Cirelli, G.~Giesen, M.~Taoso et al., work in progress. 

\bibitem{galprop}
{\sc Galprop}'s \myurl{galprop.stanford.edu}{website}.

\bibitem{dragon}
C.~Evoli, D.~Gaggero, D.~Grasso and L.~Maccione,
  JCAP {\bf 0810} (2008) 018
  [arXiv:\hhref{0807.4730} [astro-ph]].
  D.~Gaggero, L.~Maccione, G.~Di Bernardo, C.~Evoli and D.~Grasso,
  Phys.\ Rev.\ Lett.\  {\bf 111} (2013) 2,  021102
  [arXiv:\hhref{1304.6718} [astro-ph.HE]].
{\sc Dragon}'s \myurl{http://www.dragonproject.org/Home.html}{website}.  

\bibitem{Ghisellini}
  G.~Ghisellini, P.~W.~Guilbert and R.~Svensson, ApJ 334, L5 (1988).
  
\bibitem{Beck:2011he} 
  R.~Beck,
  AIP Conf.\ Proc.\  {\bf 1381}, 117 (2011)
  [arXiv:\hhref{1104.3749} [astro-ph.CO]].

\bibitem{Jansson:2009ip}
  R.~Jansson, G.~R.~Farrar, A.~H.~Waelkens and T.~A.~Ensslin,
  JCAP {\bf 0907} (2009) 021
  [arXiv:\hhref{0905.2228} [astro-ph.GA]].
  
  \bibitem{Pshirkov:2011um}
  M.~S.~Pshirkov, P.~G.~Tinyakov, P.~P.~Kronberg and K.~J.~Newton-McGee,
  Astrophys.\ J.\  {\bf 738} (2011) 192
  [arXiv:\hhref{1103.0814} [astro-ph.GA]].
  
  \bibitem{Beck:2013bxa} 
  R.~Beck and R.~Wielebinski,
 in `Planets, Stars and Stellar Systems. Volume 5: Galactic Structure and Stellar Populations', ed. T.D. Oswalt, G. Gilmore, Springer (2013),
  [arXiv:\hhref{1302.5663} [astro-ph.GA]].
  
\bibitem{Sun2008} 
  X.Sun, W.Reich, A.Waelkens, T.En\ss ling,
  A\&A {\bf 477}, 573 (2008)
  [arXiv:\hhref{0711.1572} [astro-ph]].
  
\bibitem{Jansson:2012rt} 
R.~Jansson and G.~R.~Farrar,
  Astrophys.\ J.\  {\bf 761} (2012) L11
  arXiv:\hhref{1210.7820} [astro-ph.GA].

\bibitem{Jansson:2012pc}
  R.~Jansson and G.~R.~Farrar,
  Astrophys.\ J.\  {\bf 757} (2012) 14
  [arXiv:\hhref{1204.3662} [astro-ph.GA]].

\bibitem{Sun:2010sm}
  X.~Sun and W.~Reich,
  Res.\ Astron.\ Astrophys.\  {\bf 10} (2010) 1287
  [arXiv:\hhref{1010.4394} [astro-ph.GA]].
  
\bibitem{Beck:2008ty} 
  R.~Beck,
  AIP Conf.\ Proc.\  {\bf 1085}, 83 (2009)
  [arXiv:\hhref{0810.2923} [astro-ph]].
  
\bibitem{Strong:1998fr} 
  A.~W.~Strong, I.~V.~Moskalenko and O.~Reimer,
  Astrophys.\ J.\  {\bf 537}, 763 (2000)
  [Erratum-ibid.\  {\bf 541}, 1109 (2000)]
  [\hhref{astro-ph/9811296}].
      
\bibitem{Han2006} 
  J.L.~Han,  R.N.~Manchester,  A.G.~Lyne,  G.J.~Qiao and W.~van~Straten,
 Astrophys.\ J.\  {\bf 642}, 868 (2006)
  [\hhref{astro-ph/0601357}].
  
  \bibitem{Strong:1998pw}
  A.~W.~Strong and I.~V.~Moskalenko,
  Astrophys.\ J.\  {\bf 509} (1998) 212
  [\hhref{astro-ph/9807150}].

  \bibitem{Moskalenko:1997gh}
  I.~V.~Moskalenko and A.~W.~Strong,
  Astrophys.\ J.\  {\bf 493} (1998) 694
  [\hhref{astro-ph/9710124}].
    
\bibitem{Strong:2011wd} 
  A.W.~Strong, E.~Orlando and T.R.~Jaffe,
  Astron.\ Astrophys.\  {\bf 534}, A54 (2011)
  [arXiv:\hhref{1108.4822} [astro-ph.HE]].
  
  
\bibitem{DiBernardo:2012zu}
  G.~Di Bernardo, C.~Evoli, D.~Gaggero, D.~Grasso and L.~Maccione,
  JCAP {\bf 1303} (2013) 036
  [arXiv:\hhref{1210.4546} [astro-ph.HE]].  
  
  \bibitem{Bringmann:2011py}
  T.~Bringmann, F.~Donato and R.~A.~Lineros,
  JCAP {\bf 1201} (2012) 049
  [arXiv:\hhref{1106.4821}].
  
  \bibitem{Orlando:2013ysa}
  E.~Orlando and A.~Strong,
  arXiv:\hhref{1309.2947}.
  
  \bibitem{Fornengo:2014mna}
  N.~Fornengo, R.~A.~Lineros, M.~Regis and M.~Taoso,
  JCAP {\bf 1404} (2014) 008
  [arXiv:\hhref{1402.2218} [astro-ph.CO]].
  
\bibitem{DiMauro:2014iia}
  M.~Di Mauro, F.~Donato, N.~Fornengo, R.~Lineros and A.~Vittino,
  JCAP {\bf 1404} (2014) 006
  [arXiv:\hhref{1402.0321} [astro-ph.HE]].
  
  \bibitem{Lavalle:2014kca}
  J.~Lavalle, D.~Maurin and A.~Putze,
  Phys.\ Rev.\ D {\bf 90} (2014) 8,  081301
  [arXiv:\hhref{1407.2540} [astro-ph.HE]].

  \bibitem{Fermidiffuse}  
 M.~Ackermann et al. [Fermi Collaboration], ``Fermi-LAT Observations of the Diffuse $\gamma$-Ray Emission: Implications for Cosmic Rays and the Interstellar Medium'', Astrophys. J. 750 (1), article id.~3 (2012). 
  
  \bibitem{Giesen:2015ufa}
  G.~Giesen, M.~Boudaud, Y.~Genolini, V.~Poulin, M.~Cirelli, P.~Salati, P.~D.~Serpico and J.~Feng {\it et al.},
  arXiv:\hhref{1504.04276} [astro-ph.HE].
  
  \bibitem{ISRFGalprop}
  The files are available on {\sc Galprop}'s \myurl{galprop.stanford.edu/resources.php?option=data}{website}. 
  As discussed in
  A.~E.~Vladimirov, S.~W.~Digel, G.~Johannesson, P.~F.~Michelson, I.~V.~Moskalenko, P.~L.~Nolan, E.~Orlando and T.~A.~Porter {\it et al.},
  Comput.\ Phys.\ Commun.\  {\bf 182} (2011) 1156
  [arXiv:\hhref{1008.3642} [astro-ph.HE]],
  the newest files are based on calculations using the {\sc Fr}a{\sc nkie} code (Fast Radiation transport Numerical Kode for Interstellar Emission), described in 
  T.~A.~Porter, I.~V.~Moskalenko, A.~W.~Strong, E.~Orlando and L.~Bouchet,
  Astrophys.\ J.\  {\bf 682} (2008) 400
  [arXiv:\hhref{0804.1774} [astro-ph]].
  
  \bibitem{Moskalenko:2001ya}
  I.~Moskalenko, A.~Strong, J.~Ormes, M.~Potgieter,
  Astrophys.\ J.\  {\bf 565} (2002) 280
  [\hhref{astro-ph/0106567}].
    
  \bibitem{website}
 \myurl{www.marcocirelli.net/PPPC4DMID.html}{www.marcocirelli.net/PPPC4DMID.html}

  \bibitem{Blumenthal:1970gc}
  G.~R.~Blumenthal and R.~J.~Gould,
  Rev.\ Mod.\ Phys.\  {\bf 42} (1970) 237.

\bibitem{Schlickeiser}
Reinhard Schlickeiser, `Cosmic ray astrophysics', Springer, 2002.
  
\bibitem{Evoli:2012ha}
  C.~Evoli, D.~Gaggero, D.~Grasso and L.~Maccione,
  Phys.\ Rev.\ Lett.\  {\bf 108} (2012) 211102
  [arXiv:\hhref{1203.0570} [astro-ph.HE]].

\bibitem{FermiLAT:2012aa}
  [Fermi-LAT Collaboration],
  Astrophys.\ J.\  {\bf 750} (2012) 3
  [arXiv:\hhref{1202.4039} [astro-ph.HE]].
  
  \end{thebibliography}
\end{document}